\documentclass[12pt]{article}
\usepackage{epsfig,amssymb,amsmath,psfrag}
\usepackage{cite}

\def\Reggea{\quad\mathrel{\mathop{\rm lim}\limits_{u \ll v }} 
\;\; \mathrel{\mathop{\rm lim}\limits_{u,v \ll 1 }}\quad }

\def\Reggeb{ \quad\mathrel{\mathop{\rm lim}\limits_{u,v \ll 1 }}
\;\; \mathrel{\mathop{\rm lim}\limits_{u \ll v}}\quad }
\def\logst{ {\cal L} }
\def\soverm{ {s \over m^2} }
\def\toverm{ {t \over m^2} }
\def\regge{\quad \mathrel{\mathop{\longrightarrow}\limits_{s \gg t}}\quad}
\def\eps{\epsilon}

\def\Ell{{(L)}}
\def\One{{(1)}}
\def\Two{{(2)}}
\def\Three{{(3)}}
\def\suml{\sum_{\ell=1}^\infty}
\def\pel{{(\ell)}}
\def\cG{  {\cal G}  }
\def\tc {  \tilde{c}  }
\def\arg{ (p_i, \varepsilon_i) }
\def\sumi{\sum_{i=0}^L}
\def\i{k}
\def\j{l}
\def\prodk{\prod_{\i=0}^L}
\def\prodl{\prod_{\j=1}^L}
\def\al{\alpha}
\def\bet{\beta}
\def\ga{\gamma}

\def\intxyz{ \int_0^\infty  \prodk d\al_\i \prodl d\bet_\j ~d\ga_\j }
\def\intmxyz{\int_0^\infty  \prodk (-d\al_\i) \prodl d\bet_\j ~d\ga_\j}
\def\intyz{\int_0^\infty  \prodl d\bet_\j ~d\ga_\j}
\def\ILlad{ I_{La} } 

\def\ie{{i.e.}}
\def\viz{{viz.}}
\def\eg{{e.g.}}
\def\eqn#1{eq.~(\ref{#1})} \def\Eqn#1{Equation~(\ref{#1})}
\def\eqns#1#2{eqs.~(\ref{#1}) and~(\ref{#2})}

\def \bl  {\begin{align*}}
\def \el  {\end{align*}}
\def \be  {\begin{equation}}
\def \ee  {\end{equation}}
\def \ba  {\begin{eqnarray}}
\def \ea  {\end{eqnarray}}
\def \baa {\begin{eqnarray*}}
\def \eaa {\end{eqnarray*}}
\def \bb  {\begin {thebibliography} }
\def \eb  {\end{thebibliography}}
\def \lab #1 {\label{#1}}

\def \e  {\mathop{\rm e}\nolimits}

\newcommand{\cN}{{\cal N}}
\newcommand{\cO}{{\cal O}}

\begin{document}

\begin{flushright}
HU-EP-09/64\\
BOW-PH-146\\
BRX-TH-615\\
Brown-HET-1591
\end{flushright}
%\vspace{25mm}
\vspace{3mm}

\begin{center}
{\Large\bf\sf  
Higgs-regularized three-loop four-gluon amplitude in $\cN=4$ SYM:
\\
exponentiation and Regge limits}

\vskip 5mm 
Johannes M. Henn$^{a}$, 
Stephen G. Naculich\footnote{Research supported in part by the 
NSF under grant PHY-0756518}$^{,b}$, 
Howard J. Schnitzer\footnote{Research supported in part 
by the DOE under grant DE--FG02--92ER40706}$^{,c}$
and Marcus Spradlin\footnote{Research supported in part
by the DOE under grant DE--DE-FG02-91ER40688\\
{\tt 
henn@physik.hu-berlin.de, naculich@bowdoin.edu, schnitzr@brandeis.edu, spradlin@het.brown.edu}
}$^{,d}$
\end{center}

\begin{center}
$^{a}${\em Institut f\"ur Physik\\
Humboldt-Universit\"at zu Berlin, 
Newtonstra\ss{}e15, D-12489 Berlin, Germany}

\vspace{5mm}

$^{b}${\em Department of Physics\\
Bowdoin College, Brunswick, ME 04011, USA}

\vspace{5mm}

$^{c}${\em Theoretical Physics Group\\
Martin Fisher School of Physics\\
Brandeis University, Waltham, MA 02454, USA}

\vspace{5mm}

$^{d}${\em 
Brown University, Providence, Rhode Island 02912, USA }

\end{center}

\vskip 2mm

\begin{abstract}

We compute the three-loop contribution to the $\cN=4$ supersymmetric
Yang-Mills planar four-gluon amplitude using the recently-proposed Higgs
IR regulator of Alday, Henn, Plefka, and Schuster.  In particular,
we test the proposed exponential ansatz for the four-gluon amplitude
that is the analog of the BDS ansatz in dimensional regularization.
By evaluating our results at a number of kinematic points, and also in
several kinematic limits, we establish the validity of this ansatz at
the three-loop level.

We also examine the Regge limit of the planar four-gluon amplitude
using several different IR regulators: dimensional regularization,
Higgs regularization, and a cutoff regularization.  In the latter two
schemes, it is shown that the leading logarithmic (LL) behavior of the
amplitudes, and therefore the lowest-order approximation to the gluon
Regge trajectory, can be correctly obtained from the ladder approximation
of the sum of diagrams.  In dimensional regularization, on the other hand,
there is no single dominant set of diagrams in the LL approximation.
We also compute the NLL and NNLL  behavior of the $L$-loop ladder diagram
using Higgs regularization.

\end{abstract}

\vfil\break

\section{Introduction}
\setcounter{equation}{0}

Recent years have witnessed significant advances in 
$\mathcal{N}=4$ supersymmetric Yang-Mills theory (SYM) in four dimensions.  
The AdS/CFT
conjecture relating $\mathcal{N}=4$ SYM theory to 
maximally supersymmetric string theory on AdS${}_{5}$ $\times$ S${}^{5}$ 
reveals various integrable structures.  
One aspect of recent progress is a greater understanding 
of the structure of on-shell scattering amplitudes,
at both tree level and loop level. 
For example, based on an iterative structure found at two 
loops \cite{Anastasiou:2003kj},  
Bern, Dixon and Smirnov (BDS) 
conjectured an all-loop form of the 
maximally-helicity-violating 
planar $n$-gluon amplitude \cite{Bern:2005iz}, 
which is believed to be correct for $n=4$ and $5$, 
but requires modification by a function of cross-ratios for
six or more gluons \cite{Drummond:2007cf,Alday:2007he,Bern:2008ap,Drummond:2008aq,DelDuca:2009au}.  

The simplicity of the BDS form for four- and five-gluon 
loop amplitudes arises from a hidden symmetry of 
the planar theory, \viz, dual conformal invariance \cite{Drummond:2006rz}.
Dual conformal symmetry is also present in the theory at strong coupling,
and can be seen in a string theory description
of the scattering amplitudes, 
which are identified with Wilson loop expectation values in a 
T-dual AdS space \cite{Alday:2007hr}.  
The dual conformal symmetry of scattering amplitudes is understood 
as the usual conformal symmetry of the dual Wilson loops.  
In fact, this scattering amplitude/Wilson loop relation 
extends to weak coupling as
well \cite{Drummond:2007aua,Brandhuber:2007yx,Drummond:2007cf} 
(for reviews see \cite{Alday:2008yw,Henn:2009bd}). 

The dual conformal symmetry is anomalous at loop level
due to ultraviolet divergences
associated with the cusps of the Wilson loops
(which correspond to infrared (IR) divergences of the scattering amplitudes).
Anomalous Ward identities can be derived for the Wilson loop expectation 
values \cite{Drummond:2007au},
whose solution is unique up to a function of conformal cross-ratios.
The BDS ansatz satisfies the anomalous Ward identity,
therefore it is exact for Wilson loops with four and five cusps, 
since there are no cross-ratios in these cases.  
Assuming the Wilson loop/scattering amplitude duality, 
this implies the validity of the BDS ansatz for
four- and five-gluon amplitudes.

Dual conformal symmetry extends to a dual superconformal 
symmetry \cite{Drummond:2008vq},
which is a symmetry of all tree-level amplitudes \cite{Drummond:2008vq, Brandhuber:2008pf,Drummond:2008cr}.
(This dual superconformal symmetry can also be understood from the 
string theory point of view by means of fermionic 
T-duality \cite{Berkovits:2008ic, Beisert:2008iq, Beisert:2009cs}.)
The dual superconformal symmetry combines with the conventional
superconformal symmetry of $\mathcal{N}=4$ SYM theory to form a Yangian
symmetry \cite{Drummond:2009fd}.
The IR divergences of the loop amplitudes, however,
{\it a priori}
destroy both the ordinary and dual superconformal symmetries 
(and therefore the Yangian symmetry);
this breakdown is explicitly seen in dimensional regularization.
Unlike the dual conformal symmetry, the breaking of the 
ordinary conformal symmetry is not (yet) under control
(see however refs.~\cite{Korchemsky:2009hm,Bargheer:2009qu,Sever:2009aa}
for progress in this direction),
and so the role of the Yangian symmetry for loop amplitudes is unclear.

In practice, it is desirable to use a regulator that preserves as many
symmetries as possible.
Recently, Alday, Henn, Plefka, and Schuster (AHPS) 
presented a regulator which, in contrast to dimensional regularization,
leaves the dual conformal symmetry unbroken \cite{Alday:2009zm}. 
In this approach, the SYM theory is considered on the Coulomb
branch where scalar vevs break the gauge symmetry, causing some 
of the gauge bosons to become massive through the Higgs mechanism.
Planar gluon scattering amplitudes on this branch can be computed using
scalar diagrams in which some of the internal and external states are
massive, regulating the IR divergences of the scattering amplitudes   
(for earlier references see
\cite{Alday:2007hr, Kawai:2007eg,Schabinger:2008ah,McGreevy:2008zy}).
The diagrams remain dual conformal invariant, however, provided that
the dual conformal generators are taken to act on the masses 
as well as on the kinematical variables.    
The diagrams that can appear
in the scattering amplitudes are highly constrained by the
assumption of (extended) dual conformal symmetry.
There is one point in moduli space for which all the lines along the
periphery of the diagrams have mass $m$, while the external states
and the lines in the interior of the diagram are all massless.  
This is believed to be sufficient to regulate all IR divergences of
planar scattering amplitudes.
The original SYM theory is then recovered by taking $m$ small.

Using this Higgs regulator, 
AHPS computed the $\cN=4$ SYM four-gluon amplitude at one and two loops.
They showed that the planar two-loop amplitude
satisfies an iterative relation analogous to the one that
holds in dimensional regularization \cite{Anastasiou:2003kj},
and suggested that exponentiation might extend to higher loops
in an analog of the BDS ansatz for the four-gluon amplitude:
\ba
\label{intro-bds} 
\log M(s,t) 
&=& 
-\, {1\over 8} \gamma(a) 
\left[ \log^2 \Bigl(\soverm \Bigr) + \log^2 \Bigl(\toverm\Bigr) \right]
- \tilde{\cG}_{0}(a) 
\left[ \log \Bigl(\soverm\Bigr) + \log \Bigl(\toverm\Bigr) \right]
\nonumber\\
&&+\, {1\over 8} \gamma(a) 
\left[ \log^2 \left(s\over t\right) + \pi^2\right] +{\tc}(a) + \cO(m^2) 
\ea
where $M(s,t)$ is the ratio of the all-orders planar amplitude
to the tree-level amplitude,
$\gamma(a)$ is the cusp anomalous dimension (\ref{defgamma}),
and $\tilde{\cG_0} (a)$ and $\tc (a)$ are analogs of functions
appearing in the BDS ansatz in dimensional regularization (\ref{altbds}).
Let us emphasize that the nontrivial content of the BDS ansatz is the
statement about the finite terms; the IR singular terms of the amplitude
are expected to obey \eqn{intro-bds} (or \eqn{altbds}) on general
field theory grounds.\footnote{In ref.~\cite{Mitov:2006xs} the transition 
from amplitudes in dimensional regularization to 
amplitudes where (part of) the IR divergences are regulated by (small) masses 
was investigated. 
In this way, the $\log^2 (m^2)$ and $ \log (m^2) $ terms in \eqn{intro-bds} 
can be understood as arising from a different multiplicative renormalization 
factor relative to dimensional regularization. 
It is also conceivable that by adapting the formalism of 
ref.~\cite{Mitov:2006xs} to the present case
one could show that the terms finite as $m^2 \to 0$
follow from the corresponding formula in dimensional regularization.
This would imply that the BDS ansatz can be stated 
in a scheme-independent way.}

In this paper, we explore whether \eqn{intro-bds} continues
to hold at three loops and beyond.    
We compute the three-loop four-gluon amplitude
using Higgs regularization, assuming that only integrals invariant
under (extended) dual conformal symmetry contribute.
Mellin-Barnes techniques are used to evaluate the integrals, some
parts of which are computed numerically.
We numerically evaluate the results at a number of kinematic points, 
and obtain explicit expressions in several kinematic limits
(\eg, $s=t$ and also the Regge limit $s \gg t$).
In every case, our results confirm the expected exponential ansatz
(\ref{intro-bds}) at the three-loop level.

An important difference between Higgs regularization and
dimensional regularization is that in the former, 
IR divergences take the form of logarithms of $m^2$ 
whereas in the latter,
IR divergences appear as poles in $\epsilon$,
where $D = 4-2\epsilon$.
A consequence of this is that,
provided \eqn{intro-bds} is valid, the $L$-loop amplitude in 
Higgs regularization may be computed by simply exponentiating
$\log M(s,t)$ without regard for the $\cO(m^2)$ terms
since they continue to vanish as $m \to 0$ even when multiplied
by logarithms of $m^2$.  
In contrast, the BDS ansatz in dimensional regularization 
(\ref{altbds}) specifies $\log M(s,t)$ 
up to terms that vanish as $\epsilon \to 0$.
In exponentiating $\log M(s,t)$, these neglected terms
can combine with the IR poles to give rather complicated 
contributions to the IR-finite $L$-loop amplitude.   

To put the matter the other way around, in order to test \eqn{intro-bds}
one need not compute any $\cO(m^2)$
terms of the Higgs-regulated $L$-loop amplitudes
because they cannot make any 
contribution to the IR-finite part of $\log M(s,t)$, 
whereas to test the BDS ansatz in dimensional regularization,
one must compute $\cO(\eps)$ and higher terms in the lower-loop
amplitudes to obtain all the IR-finite contributions to $\log M(s,t)$.
This is one of several significant advantages that Higgs-regulated
amplitudes have over their dimensionally-regulated counterparts.

A second major focus of this paper is the Regge behavior 
of planar $\mathcal{N}=4$ SYM amplitudes 
using Higgs regularization.
One motivation for this is that it presents a different way 
of examining the iterative structure of the theory, 
in which some results can be obtained to all orders.  
\Eqn{intro-bds} may be rewritten as
\be
\label{intro-regge}
\log M  (s,t)
=
\left[
-  
{1 \over 4} \gamma(a) 
\log \Bigl( \toverm \Bigr)
- \tilde{\cG}_{0} (a)\right] \log\Bigl(\soverm\Bigr)
- \tilde{\cG}_{0} (a) \log \Bigl( \toverm \Bigr)
+ 
{\pi^2 \over 8} \gamma(a)
+ \tilde{c}(a)  + \cO(m^2)
\ee
in which the $\log^2 (s/m^2)$ terms have
cancelled\footnote{A similar cancellation 
occurs in the Regge limit using dimensional 
regularization \cite{Drummond:2007aua, Naculich:2007ub,
DelDuca:2008pj,Naculich:2009cv}.
Subleading-color corrections to the Regge trajectory
in dimensional regularization were also considered 
in ref.~\cite{Naculich:2009cv}.}
to leave a single $\log(s/m^2)$.
Consequently, $M(s,t)$ exhibits exact Regge behavior
\be
M(s,t) = M(t,s) = \beta (t) \left( \soverm\right)^{\alpha(t)-1}
\ee
where the all-loop-orders Regge trajectory is
\be
\label{intro-alpha}
\alpha(t) -1
= - {1 \over 4} \gamma(a) \log \Bigl(\toverm\Bigr) - \tilde{\cG}_{0} (a).
\ee
The lowest order term of the trajectory,
$- a \log (t/m^2)$,
gives rise to the leading log (LL) behavior of the $L$-loop amplitude 
\be
\label{intro-LL} 
M^\Ell (s,t) \regge   {(-1)^L \over L!} 
\log^L  \Bigl( \toverm \Bigr)
\log^L \Bigl(\soverm\Bigr) 
\ee
in the Regge limit, whereas higher-order terms in the cusp
anomalous dimension contribute to the NLL (next-to-leading-log), 
NNLL, etc. pieces
of the amplitude (see \eqn{ll-expansion}).     

There is a subtle question of order of limits that appears when considering the Regge limit.
In fact there are two ways this limit can be taken.
The first possibility (a) consists in taking the limit $m^2 \ll s,t$ first, 
and then taking the limit 
$s \gg t$. 
This order is implicit in  \eqn{intro-regge}. 
The second possibility (b) consists in taking the 
limit $s \gg t , m^2$ first, and then taking
$m^2 \ll t$. 
{\it A priori}, it is not clear that the result
will not depend on the way the limit is taken. 
Explicit calculation shows that no such ambiguity arises
for the integrals appearing at one and two loops. 
At three loops, as we will show, 
the individual integrals give different contributions in the two limits; 
the three-loop amplitude $M^\Three(s,t)$, however, is unchanged. 

It seems that both ways of taking the limit can be justified, 
with slightly different interpretations.\footnote{There 
is an analogous question in dimensional 
regularization of whether to take $\epsilon \to 0$ or $s \gg t$ 
first \cite{DelDuca:2008jg}, 
but see the Erratum in {\bf v5} \cite{DelDucaErr}.}
On the one hand, 
the Regge (a) limit seems more appropriate 
in order to make contact with results
in the massless theory.
On the other hand, the Regge (b) limit is natural 
for scattering amplitudes with masses (see 
ref.~\cite{Eden} and sec.~\ref{sec-bhabha} below).

We show in this paper that the leading log behavior (\ref{intro-LL})
stems entirely from a single scalar diagram, 
the vertical ladder,  in
the Regge (b) limit of 
the Higgs-regularized loop expansion.     
(This contrasts with dimensional regularization,
in which ladder diagrams do not dominate in the Regge limit;
the leading log behavior of the $L$-loop amplitude
receives contributions from most
of the contributing diagrams.
This might have significant implications for recent discussions
of multi-Regge behavior \cite{
Brower:2008nm,Bartels:2008ce,Bartels:2008sc,Brower:2008ia,DelDuca:2008jg,Schabinger:2009bb}.)
We also compute the NLL  and 
NNLL terms of the $L$-loop vertical ladder diagram.
Other Higgs-regularized diagrams also contribute 
to the NLL, NNLL, etc. terms of the amplitude,
although we do not yet have an all-orders characterization 
of which diagrams contribute
to each order in the leading log expansion.

In ref.~\cite{Schnitzer:2007rn},
the LL approximation to the gluon Regge trajectory 
was computed using a ladder approximation and an alternative IR regulator,
in which the external lines are massless
while all internal lines of the diagrams 
are given a common mass (which cuts off the IR divergences)\footnote{This 
procedure is closely related to an off-shell regulator.  
If all physical states have a common mass $m$ arising from
a conventional Higgs mechanism, then external lines 
with $p^2 =0$ are off-shell from this point of view.}.
In the LL limit, this ``cutoff regulator'' yields results identical
to those found in this paper.\footnote{The
Higgs regulator considered in this paper does
not seem directly applicable to computing 
the subleading-color contributions of $\cN=4$ SYM amplitudes, 
which are not dual-conformal-invariant,
and which involve non-planar diagrams.
The cutoff regulator might be more suitable in this respect.}
The reason for this, as we will see, is that
the LL approximation to the vertical ladder
is insensitive to the masses of the propagators constituting
the rungs of the ladder,
and this is the only difference between the cutoff and Higgs
regulators.
We also compute the NLL contribution to the
$L$-loop vertical ladder diagram using 
the cutoff regulator, which differs from the Higgs regulator
by scheme-dependent constants.

The Regge (b) limit of the scattering amplitudes
can be understood by performing the Higgs regularization 
at a different point on the Coulomb branch of the theory.
In this case, the scattered particles are massive,
whereas some of the internal lines remain massless,
eliminating the collinear but not the soft IR divergences.
We suggest
how the Regge behavior of the four-point amplitude
can be understood from the cusp anomalous dimension
of a Wilson line with a non-light-like cusp.

The paper is organized as follows:
In sec.~\ref{sect-revdcs}, we give a short review of dual conformal symmetry.
Exponentiation of the planar four-gluon amplitude 
in dimensional regularization and in Higgs regularization 
are discussed in section \ref{sect-exp},
and the three-loop amplitude in Higgs regularization is 
computed.
The Regge limit of the planar four-gluon amplitude 
in several regularization schemes is examined in sec.~\ref{sect-regge},
and the first three terms in the leading log expansion of 
the $L$-loop vertical ladder diagram are computed.
Most of the technical details are relegated to three appendices.  

\section{Short review of dual conformal symmetry}
\setcounter{equation}{0}
\label{sect-revdcs}

In this section we give a short review of dual conformal symmetry,
with a particular focus on four-gluon amplitudes.  Hints for dual
conformal symmetry first appeared as an observation that the loop
integrals contributing to planar four-gluon scattering amplitudes in
$\mathcal{N}=4$ SYM theory have special properties when written in a dual
coordinate space \cite{Drummond:2006rz}.

\begin{figure}[t]
 \psfrag{x1}[cc][cc]{$x_{1}$}
\psfrag{x2}[cc][cc]{$x_{2}$}
\psfrag{x3}[cc][cc]{$x_{3}$}
\psfrag{x4}[cc][cc]{$x_{4}$}
\psfrag{xa}[cc][cc]{$x_{a}$}
\psfrag{p1}[cc][cc]{$p_{1}$}
\psfrag{p2}[cc][cc]{$p_{2}$}
\psfrag{p3}[cc][cc]{$p_{3}$}
\psfrag{p4}[cc][cc]{$p_{4}$}
\psfrag{k}[cc][cc]{$k$}
 \centerline{
 {\epsfxsize14cm  \epsfbox{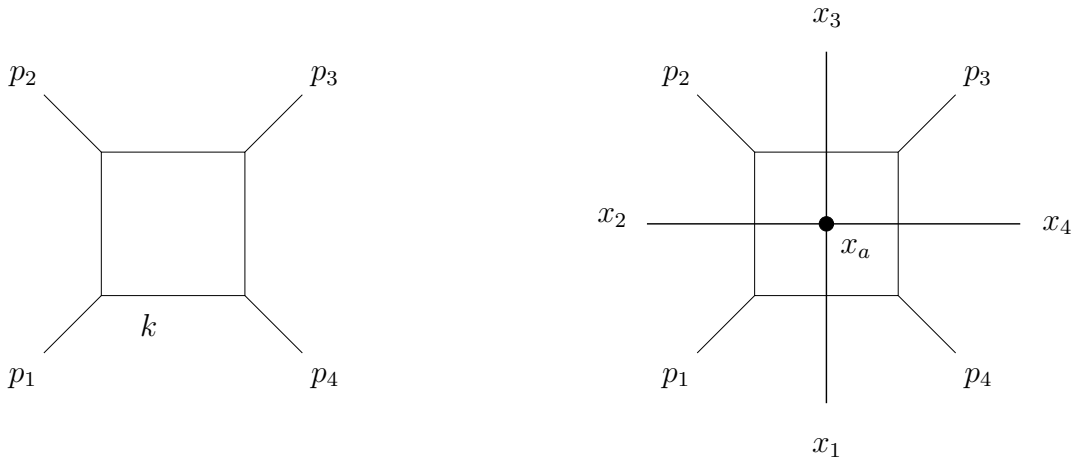}}
} 
\caption{\small
One-loop scalar box diagram representing the on-shell integral 
(\ref{int-1loop}) 
together with its dual diagram (\ref{int-1loopdual}). 
The numerator factor $(p_{1}+p_{2})^2 (p_{2}+p_{3})^2 = x_{13}^2 x_{24}^2$ 
is not displayed.}
\label{fig-oneloop4pt}
\end{figure}

Let us recall that the full four-gluon amplitude can be decomposed in a
trace basis, the coefficients of which are referred to as color-ordered
amplitudes \cite{Mangano:1987xk,Berends:1987cv,Mangano:1988kk}.
The color-ordered 
planar  (\ie, large $N$)      
amplitudes may be written in a loop expansion
\be
A \arg = \sum_{L=0}^\infty a^L   A^{(L)}  \arg
\ee
in the 't Hooft parameter \cite{Bern:2005iz}
\be
a \equiv {g^2 N \over 8 \pi^2} \left( 4 \pi \e^{-\gamma_{\rm E} } \right)^\eps 
\ee
where $\gamma_{\rm E}$ is Euler's constant, and 
$\eps = {1 \over 2} (4-D)$ 
with $\eps<0$ to regulate IR divergences.
In $\cN=4$ SYM theory, loop corrections to the four-gluon amplitude 
have the same helicity dependence as the tree amplitude, 
so we factor out the tree amplitude 
to express the amplitude as a function $M(s,t)$ of the kinematic variables 
$s= (p_1+p_2)^2$ and $t = (p_2 + p_3)^2$ only\footnote{\label{metric}In 
this paper,
we follow the $-+++$ metric conventions of ref.~\cite{Alday:2009zm},
so that $s$ is negative for positive CM energy.
The amplitude $M(s,t)$ will be real for $s$ and $t$ both positive. }
\be
\label{allloops}
 A \arg \equiv A^{(0)} \arg  \, M(s,t) ,
\qquad
M (s,t) = 1 + \sum_{L=1}^\infty a^L M^\Ell (s,t) \,.
\ee
At one loop, we have 
\be
\label{oneloopM}
M^\One(s,t)  = -  \frac{1}{2} I_{1}  (s,t)
\ee
with the one-loop integral $I_{1}$ given by
\begin{equation}
\label{int-1loop}
I_{1} (s,t) 
=  \left( e^{\gamma_{\rm E} } \mu^2 \right)^{\epsilon} \, 
\int \frac{d^{D}k}{i \pi^{D/2}} 
\frac{ (p_1 + p_2 )^2 (p_2 + p_3 )^2 }{k^2 (k+p_1 )^2 
(k+p_1 +p_2 )^2 (k-p_4)^2}\,, 
\end{equation}
where the external states are on-shell: $p_{i}^2=0$. 
After a change of variables to a dual coordinate 
space \cite{Drummond:2006rz,Broadhurst:1993ib},
\begin{equation}\label{dual-variables}
p_{i}^{\mu} = x^{\mu}_{i} - x_{i+1}^{\mu}\,, \qquad x_{5} \equiv x_{1}\,,
\end{equation}
one obtains
\begin{equation}\label{int-1loopdual}
I_{1} (s,t) =  \left( e^{\gamma_{\rm E} } \mu^2 \right)^{\epsilon}\, 
\int \frac{d^{D}x_{a}}{i \pi^{D/2}} 
\frac{ x_{13}^2  x_{24}^2 }{x_{1a}^2 x_{2a}^2 x_{3a}^2 x_{4a}^2}\,,
\end{equation}
where now the on-shell conditions read $x_{i,i+1}^2=0$.
Note that this change of variables can be most easily done in a graphical way 
(see fig.~\ref{fig-oneloop4pt}).
{}From \eqn{int-1loopdual} one can 
see that for $D=4$ the integral would be invariant under conformal transformations in the dual coordinate space \cite{Drummond:2006rz,Broadhurst:1993ib};
hence the name ``dual conformal symmetry.''
Due to infrared divergences one cannot set $D=4$, and hence the 
aforementioned symmetry is broken,
which is why such integrals were later called ``pseudoconformal.''
It was found that at least up to four loops all integrals appearing in the four-particle scattering amplitude have this property \cite{Bern:2006ew,Drummond:2007aua}. 
All this hinted at some deeper underlying structure.

{}From the practical point of view, assuming that only pseudoconformal
integrals contribute to an amplitude proved to be a useful 
guiding principle\footnote{Note that there can 
be subtleties 
about which 
integrals should be called pseudoconformal
(see \eg~ref.~\cite{Bern:2008ap}),
having to do with the peculiarities of dimensional regularization/reduction.}
(see \eg~refs.~\cite{Bern:2007ct,Bern:2008ap}).
On the other hand, the inevitable breaking of the symmetry for $D\neq 4$
was not under control.
This changed when it was realized that the (finite part of the) 
logarithm of the amplitude satisfies certain
anomalous Ward identities, which were initially derived for Wilson loops
and are conjectured to hold for scattering amplitudes as well. Assuming
the Ward identities hold for the scattering amplitudes, they explain
the correctness of the BDS ansatz for four and five scattered 
particles \cite{Drummond:2007cf, Drummond:2007au}.

Initially dual conformal symmetry could be
applied to the $n$-gluon amplitude $M_{n}$ 
for maximally-helicity-violating (MHV) amplitudes only, 
as can be seen from the fact that the variables in
\eqn{dual-variables} do not carry helicity.
In ref.~\cite{Drummond:2008vq}, it was shown how to incorporate the helicity
information and to define dual (super)conformal symmetry for arbitrary
amplitudes, both MHV and non-MHV.  
The predictions of ref.~\cite{Drummond:2008vq}
about how dual conformal symmetry is realized at tree and loop level 
(through an anomalous Ward identity) have by now been checked to one-loop 
order \cite{Drummond:2008bq,Brandhuber:2009xz,Elvang:2009ya,Brandhuber:2009kh}.
The status of dual {\it super}conformal symmetry at loop level, which
is related to the conventional superconformal symmetry, is under
investigation \cite{Korchemsky:2009hm,Bargheer:2009qu,Sever:2009aa}.

The symmetries mentioned above can also be seen at strong coupling using the AdS/CFT correspondence.
In a groundbreaking paper \cite{Alday:2007hr}, a prescription for computing scattering amplitudes at strong coupling was given.
There, a bosonic T-duality was used that maps the original AdS space to a dual AdS space. Dual conformal
symmetry can then be identified with the isometries of the dual AdS space (up to the issue of regularization).
It was later shown that the 
bosonic T-duality can be supplemented with a fermionic 
T-duality \cite{Berkovits:2008ic, Beisert:2008iq, Beisert:2009cs},
which leads to the counterpart of the dual superconformal symmetry found in the field theory.
The analysis of ref.~\cite{Berkovits:2008ic} is valid to all orders 
in the gauge coupling constant;
however, just as in the field theory, 
introducing a regulator may break the symmetry. It would be interesting
to address this question in the string theory 
approach. 
A somewhat related question is how the helicity dependence
of the scattering amplitudes is encoded in the string theory setup.
(At strong coupling, it is argued to be an overall factor, which can
be ignored, but certainly this is not the case at lower orders in the
coupling constant.)

In a recent paper \cite{Alday:2009zm}, a new regularization of planar amplitudes inspired by the string theory setup of 
refs.~\cite{Alday:2007hr,Berkovits:2008ic} was advocated (see also 
refs.~\cite{Kawai:2007eg,Schabinger:2008ah,McGreevy:2008zy}). 
In the string theory, besides the usual stack of $N$ branes at  $z=0$
($z$ being the radial AdS coordinate, with the AdS radius normalized to unity), 
$M$ further branes are placed 
at distances $z_{i}=1/m_{i}$. 
The scattering takes place on the stack of $M$ branes.
In the field theory, this 
corresponds to 
going to the Coulomb branch of $\mathcal{N}=4$ SYM. Specifically,
one starts with a gauge group $U(N+M)$ and breaks it to 
$U(N) \times U(M) \to U(N) \times U(1)^M$.
This leads to masses $| m_{i_1 } - m_{i_2} | $ for fields with labels in the $U(M)$, masses $| m_{i} |$ for fields with
mixed $N,M$ gauge labels, while the $U(N)$ fields remain massless. 
We then consider the scattering of fields with indices in the $U(M)$ 
part of the gauge group and take $N \gg M$.
(In other words, we drop all diagrams containing loops of particles with
gauge indices $M$.)
Therefore, the internal labels will all be in the $N$ part of the gauge group, 
while the particles running along the perimeter of all Feynman diagrams will have mixed $(M,N)$ labels, and hence will
have massive propagators. The latter make the integrals IR finite, and there is no need to use dimensional regularization.
Of course, strictly speaking one is considering a different theory, but the original theory is approached in the 
small mass limit.\footnote{Unfortunately, 
since the requirement of finiteness imposes keeping $N \gg M$, one cannot reproduce the
non-planar scattering amplitudes of the original theory.}

The string theory setup suggests that the planar amplitudes defined in this way should have an exact, \ie~unbroken,
dual conformal symmetry. This is possible because there are now additional terms in the dual conformal
generators that act on the Higgs masses. These additional terms come from the isometries of dual AdS space.
At one loop, the integral (\ref{int-1loopdual})  
is replaced by
\begin{equation}\label{int-1loopext}
I_{1} (s,t; m_1,m_2,m_3,m_4) 
= \int  \frac{d^{4}x_{a}}{i \pi^2} \frac{( x_{13}^2 +(m_1 -m_3 )^2 ) ( x_{24}^2 +(m_2 -m_4 )^2 ) } {(x_{1a}^2 +m_{1}^2) (x_{2a}^2 +m_{2}^2) (x_{3a}^2 +m_{3}^2) (x_{4a}^2 +m_{4}^2) }\,,
\end{equation}
now subject to the on-shell conditions $x_{i,i+1}^2 = - (m_i - m_{i+1} )^2$, with the identification 
$m_{5} \equiv m_{1}$.  
As anticipated, $I_{1}$ is annihilated by the extended form of 
dual conformal transformations,\footnote{This can be seen most easily
by thinking of the masses $m_{i}$ as a fifth coordinate of the dual
coordinates, defining $\hat{x}^{M}_{i} \equiv (x^{\mu}_{i} , m_{i} )$,
and considering conformal inversions in this five-dimensional space. For
further details, see ref.~\cite{Alday:2009zm}.}
\begin{equation}\label{dcs-oneloop}
\hat{K}^{\mu} \, I_{1} = 0\,,
\end{equation}
where
\begin{equation}
\hat{K}_{\mu}
=  \sum_{i=1}^{4} 
 \left[  2 x_{i}{}_{\mu} \left( x_{i}^{\nu} \frac{\partial}{\partial x_{i}^{\nu}} + m_{i} \frac{\partial}{\partial m_{i}} \right) - (x_{i}^2 +m_{i}^2) \frac{\partial}{\partial x_{i}^{\mu}} \right]   \,.
\end{equation}
Note that the integral is finite
(for $m_i \neq 0$), 
and the symmetry is exact (\ie~unbroken), 
hence there is no anomaly term on the r.h.s. of \eqn{dcs-oneloop}.
{}From \eqn{dcs-oneloop} one can deduce that the functional dependence
of $I_{1}$ is  \cite{Alday:2009zm}
\be
I_{1}(s,t;m_{1},m_{2},m_{3},m_{4}) = f(u,v),
\ee
where
\be
\label{defuv}
u = \frac{m_{1} m_{3}}{s+ (m_1 -m_3 )^2 }
\qquad{\rm and} \qquad v = \frac{m_{2} m_{4}}{t+  (m_2 -m_4 )^2} \,.
\ee
Similar restrictions hold for a higher number of external legs.
This form of dual conformal invariance in the field theory was checked for a particular  four-scalar  amplitude  at one loop \cite{Alday:2009zm}.
There it was also shown that assuming this symmetry at two loops leads to 
an iterative relation similar to that which holds in dimensional regularization.

\begin{figure}[t]
\psfrag{m1}[cc][cc]{$m_{1}$}
\psfrag{m2}[cc][cc]{$m_{2}$}
\psfrag{m3}[cc][cc]{$m_{3}$}
\psfrag{m4}[cc][cc]{$m_{4}$}
\psfrag{mn}[cc][cc]{$m_{n}$}
\psfrag{p1sq}[cc][cc]{$p_{1}^2 = -(m_1 - m_2 )^2$}
\psfrag{p2sq}[cc][cc]{$p_{2}^2 = -(m_2 - m_3 )^2$}
\psfrag{p3sq}[cc][cc]{$p_{3}^2 = -(m_3 - m_4 )^2$}
\psfrag{pnsq}[cc][cc]{$p_{n}^2 = -(m_n - m_1 )^2$}
\psfrag{dots}[cc][cc]{$\dots$}
\centerline{
{\epsfxsize11cm  \epsfbox{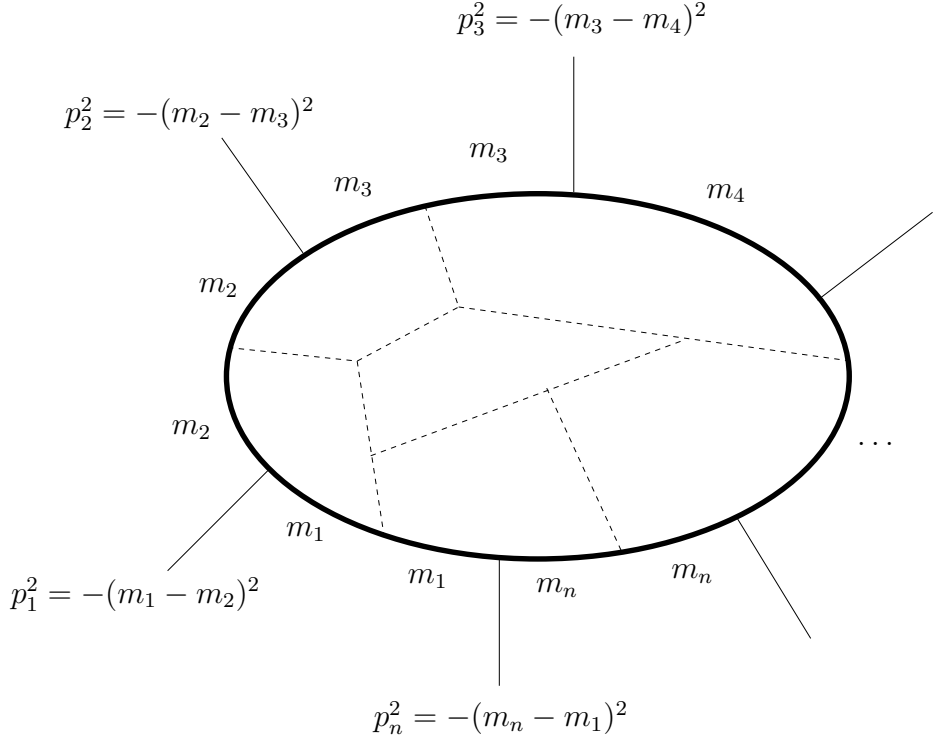}}
}
\caption{\small
Mass assignment of a generic planar diagram in the Higgs setup.  The external lines correspond to on-shell
particles with
$p_{i}^2 = -(m_{i}-m_{i+1})^2$. 
The internal dashed lines correspond to massless particles.
We call particles of mass $ | m_{i} - m_{i+1} |$ `light' and particles of mass $m_{i}$ `heavy' since the former become
massless when we consider the equal mass case $m_{i}=m$.}
\label{fig-genmasses}
\end{figure}

In fig.~\ref{fig-genmasses}, we illustrate
where the Higgs masses appear in a generic $n$-point integral in this setup. 
Thanks to dual conformal symmetry, 
we can set all masses equal without loss of generality,  
in which case all external particles are massless, 
the particles in the outer loop of a diagram are massive, 
and all particles on the inside are massless.

One can  
in principle write down all dual conformal integrals at a given loop order 
and for a given number $n$ of external legs.
The amplitude would then be given by 
a linear combination of these integrals \cite{Drummond:2007aua}
\begin{equation}\label{dcs-conjecture}
M_n = 1 +  \sum_{  \mathcal{I}  } a^{L(\mathcal{I})} \, c( \mathcal{I} ) \, \mathcal{I} \,,
\end{equation}
where the sum runs over all dual conformal 
integrals\footnote{Here we mean 
dual conformal integrals in the sense of the Higgs regulator 
of ref.~\cite{Alday:2009zm}, not the off-shell 
regulator 
of ref.~\cite{Drummond:2007aua}.} 
$\mathcal{I}$, 
with $ L(\mathcal{I})$ the loop order 
of the integral, 
and $c( \mathcal{I} )$ a (rational) coefficient
(a number in the case of MHV amplitudes). 
Further restrictions result from the 
requirement that the diagrams must arise from a scattering process;
for example, at two loops, the one-loop integral (\ref{int-1loopext}) 
cannot appear squared.
Moreover, there are restrictions on the number of propagators and numerator factors.
Finally, integrals that would be formally dual conformal invariant but are divergent despite
the introduction of the Higgs regulator should not appear \cite{Drummond:2007aua}.

\section{Exponentiation of the four-gluon amplitude}
\setcounter{equation}{0}
\label{sect-exp}

In this section, we recall the exponentiation of the 
four-point amplitude of $\cN=4$ SYM theory, 
first using dimensional regularization to regulate the
IR divergences, and then using the Higgs regularization
described in sec.~\ref{sect-revdcs}.

\subsection{Exponentiation in dimensional regularization}

On the basis of ref.~\cite{Anastasiou:2003kj}, 
Bern, Dixon, and Smirnov 
conjectured \cite{Bern:2005iz} 
that the dimensionally-regularized all-loop orders amplitude 
(\ref{allloops})
satisfies 
\be
\label{bds}
\log M(s,t) = \suml  a^\ell   
\left[ f^\pel (\eps) M^\One (s,t; \ell \eps) + C^\pel + \cO(\eps) \right]
\ee
where  $ M^\One (s,t) $ is given by \eqn{oneloopM}, 
$\eps = {1 \over 2}(4-D)$, and
\be
f^\pel (\eps)  
=
{1\over 4} \gamma^\pel   
+ {1\over 2} \eps \: \ell \: \cG_0^\pel 
+ \eps^2  f_2^\pel  
\ee
with the cusp 
$\gamma^\pel$    
and collinear
$\cG_0^\pel $ 
anomalous dimensions \cite{Korchemskaya:1992je}
given by\footnote{We use the notation of
ref.~\cite{Bern:2005iz}.   Note that
$\gamma(a) = 2 \, \Gamma_{\rm cusp}(a)$,
where $\Gamma_{\rm cusp}(a) $ is also widely used in the literature.
} 
\ba
\label{defgamma}
\gamma(a) 
&=& 
\sum_{\ell=1}^\infty a^\ell  
\gamma^\pel 
= 
4a - 4\zeta_2 a^2 + 22 \zeta_4 a^3 + \cdots
\\
\label{defG}
\cG_0(a)
&=& 
\sum_{\ell=1}^\infty  a^\ell  \cG_0^\pel =
    -  \zeta_3 a^2 + \left(4 \zeta_5 + \frac{10}{3} \zeta_2 \zeta_3 \right)  a^3 + \cdots
\ea
The constants $C^\pel$ and $f_2^\pel$ 
and the $\cO(\eps)$ terms in \eqn{bds} are not known {\it a priori}.
The BDS ansatz (\ref{bds}) may be re-expressed as
\be
\label{altbds}
\log M(s,t) =
\suml   a^\ell
\left[ 
-\frac{\gamma^\pel}{ 4 (\ell \eps)^2} 
-\frac{\cG_0^\pel} {2 \ell \eps} 
\right]
\left[
\left( \mu^2 \over s\right)^{\ell \eps}
+
\left( \mu^2 \over t\right)^{\ell \eps}
\right]
+ 
{\gamma(a)\over 8}   
\left[ \log^2 \left(s\over t\right) + \frac{4}{3}\pi^2\right] 
+ c(a)
+ \cO(\eps) \qquad
\ee
where \cite{Bern:2005iz}
\be
c(a) 
= 
\suml   a^\ell c^\pel 
= 
\suml   a^\ell
\left[ 
-\frac{2 f_2^\pel}{ \ell^2} + C^\pel
\right]
=
- \frac{\pi^4}{120} a^2 + 
\left( \frac{341}{216} \zeta_6 - \frac{17}{9} \zeta_3^2 \right)  a^3 + \cdots
\ee
Overlapping soft and collinear IR divergences
are responsible for the $1/\eps^2$ pole in \eqn{altbds}. 

While the BDS ansatz (\ref{altbds}) implies that the IR-finite
part of the {\it logarithm} of the amplitude is simply expressible
in terms of $\log (s/\mu^2)$ and $\log (t/\mu^2)$ and a set of constants
$\gamma^\pel$, $\cG_0^\pel $, and $c^\pel$,
the same is not true of the $L$-loop amplitudes $M^\Ell (s,t)$ themselves.
For example, \eqn{altbds} implies \cite{Anastasiou:2003kj}
\ba
\label{twoloop}
M^\Two (s,t)
&=&
{1 \over 2} \left[ M^\One (s,t) \right]^2 
- \frac{\gamma^\Two}{ 8 \eps^2} 
- \frac{\cG_0^\Two} {2 \eps}
+ \frac{\gamma^\Two}{ 8 \eps} 
\left[ \log \Bigl( {s \over \mu^2} \Bigr) +
\log \Bigl( {t \over \mu^2} \Bigr) \right]
\\
&+&
\frac{\cG_0^\Two} {2} 
\left[ \log \Bigl( {s \over \mu^2} \Bigr) +
\log \Bigl( {t \over \mu^2} \Bigr) \right]
- \frac{\gamma^\Two}{4}
\log \Bigl( {s \over \mu^2} \Bigr)
\log \Bigl( {t \over \mu^2} \Bigr)
+ \frac{\pi^2}{6} \gamma^\Two 
+ c^\Two 
+\cO(\eps) \,.
\nonumber
\ea
Because of interference
between the positive and negative powers of $\eps$ in 
$\left[M^\One(s,t)\right]^2$,
the 
$\cO(\eps^{-1})$ and
$\cO(\eps^0) $ terms \cite{Smirnov:1999gc} in $M^\Two(s,t) $ 
depend on more complicated functions (polylogarithms) 
of $s$ and $t$, 
which are present in the $\cO(\eps)$ and $\cO(\eps^2)$
terms \cite{Anastasiou:2003kj} of $M^\One(s,t)$.
In general $M^\Ell (s,t)$
will receive contributions from the 
coefficients of positive powers of $\epsilon$ in all lower-loop amplitudes.  

\subsection{Exponentiation in Higgs regularization}
\label{sect-exphiggs}

The Higgs mechanism reviewed in section \ref{sect-revdcs} can be used as 
a gauge-invariant regulator of the IR divergences in the 
planar massless theory, 
with the IR divergences appearing as $\log(m^2)$ terms in the amplitude, 
and any terms that vanish as $m \to 0$ are dropped.
One could ask whether, when regulated in this way,
the four-point loop amplitude satisfies iterative relations
similar to the BDS ansatz for the dimensionally-regulated amplitude.
In ref.~\cite{Alday:2009zm}, it was suggested that the analog
of \eqn{altbds} is
\ba 
\label{all-loop-higgs}
\log M(s,t) 
&=& 
-\, {1\over 8} \gamma(a) 
\left[ \log^2 \Bigl(\soverm \Bigr) + \log^2 \Bigl(\toverm\Bigr) \right]
- \tilde{\cG}_{0}(a) 
\left[ \log \Bigl(\soverm\Bigr) + \log \Bigl(\toverm\Bigr) \right]
\nonumber\\
&&+\, {1\over 8} \gamma(a) 
\left[ \log^2 \left(s\over t\right) + \pi^2\right] +{\tc}(a) + \cO(m^2) 
\ea
where $\gamma(a)$ is the cusp anomalous dimension (\ref{defgamma}),
and 
$\tilde{\cG}_{0} (a)$ and $\tc(a)$ are the analogs of 
$\cG_{0} (a)$ and $c(a)$,
but need not be the same functions
since they are scheme-dependent \cite{Alday:2009zm}.
Overlapping soft and collinear IR 
divergences\footnote{In sec.~\ref{sec-bhabha}, 
we will see that only soft divergences contribute in the Regge limit, 
giving a single logarithmic divergence.}
are responsible for the double logarithms in \eqn{all-loop-higgs}.

The $L$-loop amplitudes
$M^\Ell (s,t)$ are obtained by exponentiating \eqn{all-loop-higgs}.
In contrast to dimensional regularization, 
there is no interference between the IR-divergent $\log(m^2)$ terms
and the $\cO(m^2)$ terms in \eqn{all-loop-higgs}
since such terms vanish for $m \to 0$
order by order in the coupling constant. 
Hence the amplitudes
are simply expressed in terms of products of $\log(s/m^2)$ and
$\log(t/m^2)$ and a set of constants
$\gamma^\pel$, $\tilde{\cG}_0^\pel $, and $\tc^\pel$.
For comparison with later calculations,
we explicitly write the predictions for
the first few loop amplitudes, with $u\equiv m^2/s$ and $v\equiv m^2/t$,
\ba
\label{fullMone}
M^\One 
&=& 
- \log (v)\log(u) +\frac{1}{2} \pi ^2  + \cO(m^2)\,,
\\
M^\Two 
&=& 
{1\over 2} \log^2  (v)   \log^2(u) 
-\left(
\frac{1}{2} \pi ^2  
+
\frac{1}{4} \gamma^\Two 
\right)
\log (v)  \log(u)
 \nonumber \\
&+&
\tilde{\cG}_0^\Two  
\left[ \log(u) + \log (v) \right]
+
\left( 
\frac{1}{8} \pi ^4
+\frac{1}{8} \pi ^2 \gamma^\Two
+\tc^\Two 
\right)
 + \cO(m^2)\,,  
\label{fullMtwo}  \\
M^\Three
&=& 
 -{1\over 6} \log^3 (v) \log^3 (u) + \left( 
\frac{1}{4} \pi ^2 
+\frac{1}{4} \gamma^\Two 
\right) \log ^2(v)  \log^2 (u)
\nonumber
\\
&-& \tilde{\cG}_0^\Two
\left[  \log^2 (v)  \log (u) +  \log (v)  \log ^2(u) \right]
\nonumber\\
&-&
\left( 
\frac{1}{8} \pi ^4 
+\frac{1}{4} \pi ^2 \gamma^\Two 
+\tc^\Two  
+\frac{1}{4} \gamma^\Three
\right) \log (v) \log (u)
\nonumber\\
&+&
\left(
\frac{1}{2} \pi ^2\tilde{\cG}_0^\Two  
+\tilde{\cG}_0^\Three 
\right) 
 \left[ \log (u)   + \log (v)  \right] \nonumber \\
&+& 
\left(
\frac{1}{48} {\pi ^6}
+\frac{1}{16} \pi ^4 \gamma^\Two
+ \frac{1}{2} \pi ^2 \tc^\Two
+\frac{1}{8} \pi ^2 \gamma^\Three
+\tc^\Three  
\right)
+ \cO(m^2) 
\,,
\label{fullMthree}
\ea
where we have explicitly set
$\gamma^\One = 4$ and $\tilde{\cG}_{0}^\One = \tc^\One = 0$.

\begin{figure}[t]
\psfrag{I2}[cc][cc]{$I_{2}$}
\psfrag{I3a}[cc][cc]{$I_{3a}$}
\psfrag{I3b}[cc][cc]{$I_{3b}$}
 \centerline{
 {\epsfxsize15cm  \epsfbox{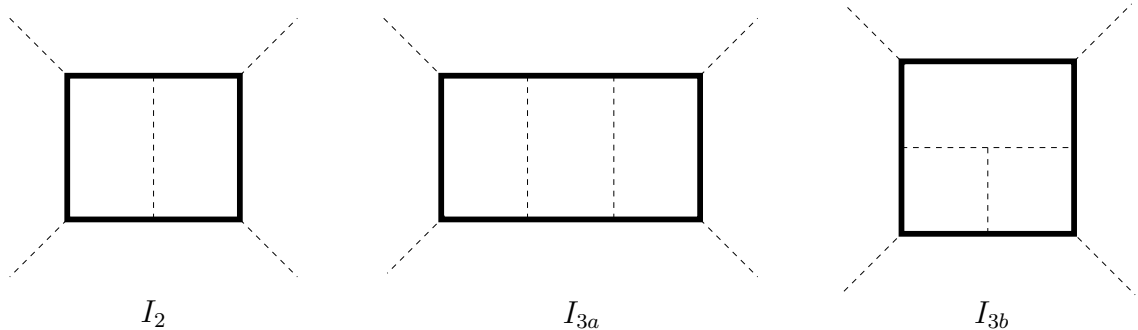}}
}
\caption{\small
Two- and three-loop four-point dual conformal diagrams. 
$I_{2}$ and $I_{3a}$ are ladder diagrams; 
$I_{3b}$ is the  
tennis-court diagram. 
Numerator factors 
(including a loop-momentum-dependent factor for the tennis court)
are omitted. 
See \eqns{defladders}{deftenniscourt}
for explicit formulae for the integrals. 
\label{fig-23loop}
}
\end{figure}

The dual-conformal integrals that contribute through two loops are 
(see figs.~\ref{fig-oneloop4pt} and \ref{fig-23loop})
\be
\label{Mhiggstwo}
M(s,t)  
= 
1 - \frac{a}{2} \, I_{1}(s,t,m^2 ) + \frac{a^2}{4} \, \left[ I_{2}(s,t,m^2 ) + I_{2}(t,s,m^2 ) \right]  
+\cO(a^3) \,.
\ee
These integrals were computed in ref.~\cite{Alday:2009zm},
and the exponential ansatz (\ref{all-loop-higgs})
was verified to two-loop order.
To determine the values of the constants in \eqns{fullMone}{fullMtwo},
it is sufficient to evaluate \eqn{Mhiggstwo} at $s=t$.
Defining $x \equiv m^2/s=m^2/t$, one finds \cite{Alday:2009zm}
\begin{eqnarray}
I_{1}(x) &=& 2 \log^2(x) - \pi^2  + \cO(x),\\
I_{2}(x) &=& \log^4(x) - \frac{2}{3} \pi^2 \log^2(x) 
- 4 \zeta_{3} \log(x) + \frac{1}{10} \pi^4 + \cO(x).
\end{eqnarray}
As discussed in ref.~\cite{Alday:2009zm}, 
these are consistent with \eqns{fullMone}{fullMtwo}
provided 
\be
\label{twoloopanom}
\gamma(a) = 4 a- 4 \zeta_2 a^2 + \cO(a^3),\qquad
\tilde{\cG}_{0} (a) = - \zeta_{3} a^2  + \cO(a^3), \qquad
\tc(a) = {\pi^4\over 120} a^2 + \cO(a^3).
\ee
The expression for $\gamma(a)$ is consistent with \eqn{defgamma}.

\begin{figure}[t]
\psfrag{I3c}[cc][cc]{$I_{3c}$}
\psfrag{I3d}[cc][cc]{$I_{3d}$}
 \centerline{
 {\epsfxsize11cm  \epsfbox{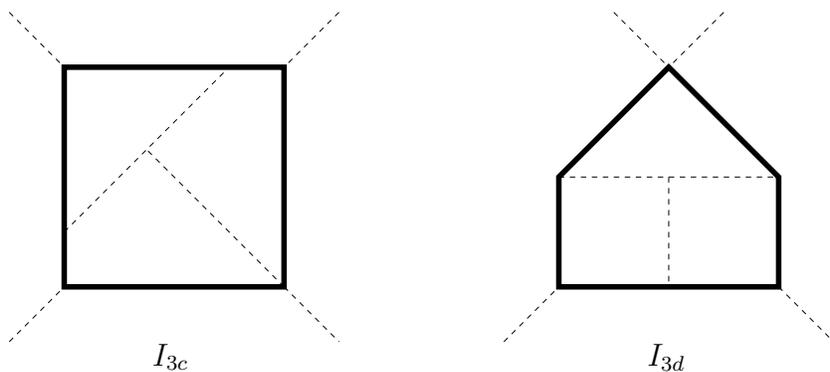}}
}
\caption{\small
Three-loop four-point dual conformal diagrams, 
with numerator factors omitted. 
Both integrals require a factor of $m^2$ in the numerator
in order to be dual-conformally-invariant.}
\label{fig-3loopcd}
\end{figure}

We now test exponentiation (\ref{all-loop-higgs})
at the three-loop level, \ie, the prediction (\ref{fullMthree}).  
Since an explicit calculation of higher-loop amplitudes using the Higgs
regulator is not (yet) available, we will start from the assumption
(\ref{dcs-conjecture}) that only dual-conformal integrals contribute.
\Eqn{dcs-conjecture} requires two ingredients: 
the set of dual-conformal integrals and their coefficients.
In order to identify 
the allowed set of dual conformal integrals it is helpful to 
use the dual notation and graphs 
introduced 
in section \ref{sect-revdcs}. 
(For more details and examples, 
see refs.~\cite{Drummond:2006rz,Bern:2006ew,Drummond:2007aua,Bern:2007ct,Nguyen:2007ya}.)

There are four dual conformal integrals that can in principle appear 
\cite{Nguyen:2007ya}
in a four-point three-loop amplitude:
$I_{3a}, I_{3b}, I_{3c}$, and $I_{3d}$. 
The first two, depicted in fig.~\ref{fig-23loop}, 
are natural dual conformal analogs of the
integrals appearing in the three-loop amplitude 
computed in dimensional regularization \cite{Bern:2005iz}.
The last two, depicted in fig.~\ref{fig-3loopcd},
are absent in dimensional regularization.
We will assume that these are the only integrals required in the
Higgs regularization, with the same coefficients 
as in the dimensional regularization result.
In general it is not valid
to take a result computed in one regularization and 
transpose it to a different regularization; 
here this procedure can be justified {\it a posteriori} 
(as we discuss below)
by imposing that the IR singular terms of the amplitude
obey the relation \eqn{all-loop-higgs} as required on general
field theory grounds.\footnote{It 
would be desirable to determine the coefficients of these
integrals using a unitarity-based method.  Indeed, once a
basis of integrals has been established, in our case using
dual conformal symmetry, (generalized) unitarity cuts are a
powerful tool to compute the coefficients of the integrals 
(see e.g. refs.~\cite{Bern:1994zx,Bern:1994cg,Buchbinder:2005wp,Bern:2007ct,Cachazo:2008dx}).
}

Given these assumptions we write
\be
\label{Mhiggsthree}
M^\Three(s,t)  
= 
- \frac{1}{8} \, \left[  I_{3a}(s,t,m^2 ) + I_{3a}(t,s,m^2 ) +2 \, I_{3b}(s,t,m^2 ) + 2 \, I_{3b}(t,s,m^2 )\right] \,.
\ee
In appendix \ref{sect-MB}, 
Mellin-Barnes (MB) representations for the integrals 
appearing in \eqn{Mhiggsthree} are derived.
The small $m^2$ limit of these MB integrals is extracted 
using the same method as in ref.~\cite{Alday:2009zm}.
At the kinematic point $s=t$, we find 
\begin{eqnarray}\label{I3a}
\
I_{3a}(x) &=& \frac{17}{90} \log^6(x) + \frac{1}{9} \pi^2 \log^4(x) -\frac{8}{3} \zeta_{3} \log^3(x) \\
&&-64.93939402 \log^2(x) 
-200.29103 \log(x)-196.597 + \cO(x)  \,, \nonumber \\
I_{3b}(x) &=& \frac{43}{180} \log^6(x) - \frac{2}{9} \pi^2 \log^4(x) - \frac{8}{3} \zeta_{3} \log^3(x) \label{I3b} \\
&&+37.8813132\, \log^2(x) 
+113.11769  \log(x)+90.0915 + \cO(x)  \,,  \nonumber
\end{eqnarray}
where the 
decimal coefficients 
are approximations obtained by numerical integration of MB integrals.
We find that the three-loop amplitude (\ref{Mhiggsthree})
at $s=t$ is consistent with the exponential ansatz (\ref{fullMthree})  
provided that 
\be
\label{threeloopanom}
\gamma^\Three \approx 23.81111114 \pm 10^{-8}, \qquad
\tilde{\cG}_{0}^{(3)} \approx 2.68887 \pm 10^{-5}, \qquad 
\tc^{(3)} \approx -9.249 \pm 10^{-3}\,.
\ee
together with \eqn{twoloopanom}.
We note that $\gamma^\Three \approx 22 \zeta_4$
is indeed the correct three-loop cusp anomalous dimension (\ref{defgamma}). 

Having obtained the coefficients from the $s=t$ case,
we are now in a position to test
the full consistency of \eqn{Mhiggsthree} with \eqn{fullMthree}.
In order to do this, we have evaluated the coefficients of 
the small $m^2$ expansion of the 
integrals $I_{3a}$ and $I_{3b}$ numerically for 
various values of the kinematical variables $s,t$.
We have found agreement with \eqn{fullMthree} within the numerical accuracy
of the calculation.

We assumed above that the coefficients of the integrals 
$I_{3c}$ and $I_{3d}$ vanish in Higgs regularization,
as they do in dimensional regularization.   
We have nevertheless evaluated these integrals in the small $m^2$ limit
using MB methods, and found that
\ba
I_{3c}(u,v)  &=&  56.23  + \cO(m^2)
\\
I_{3d}(u,v) &=&  -17.32  \log(v) - 62 + \cO(m^2)
\ea
so that, if present, they could be simply accommodated in the BDS ansatz
at three loops by 
redefining\footnote{Such a change could be detected when checking 
\eqn{all-loop-higgs} at higher orders in perturbation theory.}
$\tilde{\cG}_{0}^{(3)}$ and $\tc^{(3)}$ in \eqn{fullMthree}.
We will therefore not discuss them further in this paper.

Equation (\ref{fullMthree}) predicts the three-loop amplitude
for arbitrary values of $u$ and $v$. 
We would like to present a further test of \eqn{fullMthree} 
in the limit $s \gg t$, i.e. $u \ll v$.
This is Regge limit (a) discussed in the introduction. 
In order to find formulas for our integrals in Regge limit (a), 
we perform the small $m^2$ and subsequently the $u\ll v$ limit
in the MB integrals. 
In this way, we obtain an expression in terms of powers of  
$\log(u)$ and $\log(v)$, whose kinematic-independent coefficients are
either numbers or (relatively simple) MB integrals. Where necessary we
evaluate the latter numerically.

Let us first collect the results for the 
one- and two-loop diagrams  \cite{Alday:2009zm}. 
The one-loop box diagram gives
\be\label{I1regge}
\Reggea
I_{1}(u,v) = 
\log(u) \bigg[ 2 \log(v) + \cO(v) \bigg] 
+ 
\bigg[ - \pi^2 + \cO(v) \bigg] 
+ \cO(u) \,.
\ee
For the two-loop horizontal ladder diagram (cf. fig.~\ref{fig-23loop}),
one has 
\begin{eqnarray}\label{I2reggeu}
\Reggea
I_{2}(u,v) &=& \log(u) \left[ \frac{4}{3} \log^3(v) + \frac{4}{3} \pi^2 \log(v) + \cO(v) \right] 
\\ 
&+&  
\left[ -\frac{1}{3} \log^4(v) - 2 \pi^2  \log^2(v) 
- 4 \zeta_{3} \log(v) - \frac{7}{15} \pi^4   + \cO(v) \right] 
+\cO(u) \,,
\nonumber
\end{eqnarray}
while for the two-loop vertical ladder 
\begin{eqnarray}\label{I2reggev}
\Reggea
I_{2}(v,u) &=& 
 \log^2(u) \bigg[ 2 \log^2(v) + \cO(v) \bigg]  \nonumber \\
&+& \log(u) \left[ -\frac{4}{3} \log^3(v) -\frac{8}{3} \pi^2  \log(v) - 4 \zeta_{3}  + \cO(v) \right] 
\\
&+& 
\left[ \frac{1}{3} \log^4(v) + 2 \pi^2  \log^2(v) 
+ \frac{2}{3} \pi^4   + \cO(v) \right] 
+ \cO(u) \,. \nonumber
\end{eqnarray}
Substituting these results into \eqn{Mhiggstwo},
we find agreement with \eqns{fullMone}{fullMtwo}
with the coefficients given in \eqn{twoloopanom}.
(The limit $u \ll v$ does not affect the form 
of \eqns{fullMone}{fullMtwo}.)

We now turn to the three-loop diagrams. 
For the three-loop horizontal ladder diagram (cf. fig.~\ref{fig-23loop}),
we obtain\footnote{The coefficients of the
$\pi^4$ terms were obtained numerically,
and replaced by their probable rational equivalents.}
\begin{eqnarray}
\label{I3areggeu}
&&\hspace{-.5in}\Reggea
I_{3a}(u,v)
 = \log(u) \left[ \frac{4}{15} \log^5(v) + \frac{8}{9} \pi^2 \log^3(v) 
 + \frac{28}{45} \pi^4 \log(v)  + \cO(v) \right] \nonumber\\
&+& \left[ -\frac{7}{90} \log^6(v) -\frac{7}{9} \pi^2 \log^4(v) 
- \frac{8}{3} \zeta_{3} \log^3(v) - \frac{58}{45} \pi^4 \log^2(v) 
-35.786 \log(v) -323.7 + \cO(v) \right]
\nonumber\\
&+& \cO(u) 
\,,
\end{eqnarray}
while for the vertical three-loop ladder, we obtain
\begin{eqnarray}
&&\hspace{-0.5in}\Reggea
I_{3a}(v,u) 
~=~
\log^6 (u) \left[  \frac{1}{90} + \cO(v) \right] 
+ 
\log^5 (u) \left[ -\frac{2}{15} \log(v) + \cO(v) \right] 
\nonumber\\
&+& 
\log^4 (u) \left[ \frac{2}{3} \log^2 (v)  + \frac{2}{9} \pi^2 + \cO(v) \right] 
+
\log^3 (u) \left[ -\frac{4}{9} \log^3 (v)  -\frac{16}{9} \pi^2 \log(v) + O(v) \right] 
\nonumber\\
&+& 
\log^2 (u) \left[ 2 \pi^2 \log^2 (v)  - 8 \zeta_{3} \log(v) + \frac{44}{45} \pi^4 + \cO(v) \right] 
\\
&+& 
\log (u) \left[ \frac{2}{15} \log^5(v)  + 8 \zeta_{3} \log^2 (v)  
-\frac{74}{45} \pi^4 \log(v) + 75.717 + \cO(v) \right] \nonumber \\
&+& \left[ -\frac{2}{45} \log^6(v)  - \frac{1}{3} \pi^2 \log^4(v) 
- \frac{8}{3} \zeta_{3} \log^3 (v)  - 111.5 \log(v) + 141.2 + \cO(v) \right] + \cO(u) 
\nonumber \,.
\end{eqnarray}
For the tennis court diagram in the orientation shown in fig.~\ref{fig-23loop} we find,
\begin{eqnarray}
&&\hspace{-0.5in}\Reggea
I_{3b}(u,v) ~=~
\log^6 (u) \left[  -\frac{1}{90} + \cO(v) \right] 
+ 
\log^5 (u) \left[ \frac{2}{15} \log(v) + \cO(v) \right] 
\nonumber\\
&+&
\log^4 (u) \left[ -\frac{2}{3} \log^2 (v)  - \frac{2}{9} \pi^2 + \cO(v) \right]
+
\log^3 (u) \left[ \frac{16}{9} \log^3 (v)  +\frac{16}{9} \pi^2 \log(v) + O(v) \right] 
\nonumber\\
&+&
\log^2 (u) \left[ -\frac{4}{3}\log^4(v) -4 \pi^2 \log^2 (v)   - \frac{44}{45} \pi^4 + \cO(v) \right] 
\\
&+& 
\log (u) \left[ \frac{1}{3} \log^5(v)  + \frac{20}{9} \pi^2 \log^3 (v) 
 - 8 \zeta_{3} \log^2 (v)  +\frac{20}{9} \pi^4 \log(v) -24.8863 + \cO(v) \right] \nonumber \\
&+& \left[ \frac{1}{180} \log^6(v)   + \frac{16}{3} \zeta_{3} \log^3 (v) 
 -\frac{13}{45} \pi^4 \log^2 (v)  + 111.499 \log(v)  -206.1 + \cO(v) \right] + \cO(u) 
\nonumber \,,
\end{eqnarray}
while for the tennis court in the transposed orientation, we obtain
\begin{eqnarray}
\label{I3breggev}
&&\hspace{-0.5in}\Reggea
I_{3b}(v,u) ~=~
\log^6 (u) \left[  \frac{1}{180} + \cO(v) \right] 
 + 
\log^5 (u) \left[- \frac{1}{15} \log(v) + \cO(v) \right] 
\nonumber \\
&+& 
\log^4 (u) \left[ \frac{1}{3} \log^2 (v)  + \frac{1}{9} \pi^2 + \cO(v) \right] 
+
\log^3 (u) \left[ -\frac{8}{9} \log^3 (v)  -\frac{8}{9} \pi^2 \log(v) + O(v) \right] 
\nonumber\\
&+&  
\log^2 (u) \left[ \frac{4}{3}\log^4(v) + \frac{8}{3} \pi^2 \log^2 (v)   
+ \frac{22}{45} \pi^4 + \cO(v) \right] 
\\
&+& 
\log (u) \left[ -\frac{8}{15} \log^5(v)  - \frac{8}{3} \pi^2 \log^3 (v)  
 -\frac{8}{5} \pi^4 \log(v) + \cO(v) \right] \nonumber \\
&+& \left[ \frac{1}{18} \log^6(v)  +\frac{5}{9} \pi^2 \log^4(v) 
- \frac{8}{3} \zeta_{3} \log^3 (v)  +\frac{14}{15} \pi^4 \log^2 (v) 
 - 24.888 \log(v) +280.8 + \cO(v) \right] 
\nonumber\\
&+& \cO(u) 
\nonumber \,.
\end{eqnarray}
Plugging these into \eqn{fullMthree}, we see that the
$\log^6(u), \log^5(u)$, and $\log^4(u)$ terms cancel, 
and that the $\log^3(u)$, $\log^2(u)$, $\log(u)$, and $\log^0(u)$ terms 
of \eqn{fullMthree} are reproduced 
using the coefficients given in 
\eqns{twoloopanom}{threeloopanom}.

In summary, we used the assumption of extended dual conformal
symmetry to restrict the integrals allowed to appear in
the three-loop four-point amplitude to 
$I_{3a}, I_{3b}, I_{3c}$, and $I_{3d}$.
We then successfully checked the consistency of this assumption
with the exponential ansatz (\ref{fullMthree}). 
The first test consisted in computing the 
small $m^2$ expansion of the amplitude,
and numerically comparing to \eqn{fullMthree} for various
values of the kinematical variables $u$ and $v$. 
The second test consisted in comparing our result to \eqn{fullMthree}
in the limit $u \ll v$.

\section{Regge limit of the four-gluon amplitude}
\setcounter{equation}{0}
\label{sect-regge}

In this section, we examine the four-point amplitude $M(s,t)$
in the Regge limit $s \gg t >0$.
(Recall that $M(s,t)$ is real
in this kinematic region (see footnote \ref{metric}).
The ``physical'' Regge limit, $-s \gg t > 0$, differs
from this by an imaginary contribution.)
First we review the situation for dimensional regularization,
and then we reexamine the Regge limit using Higgs regularization
and cutoff regularization.
The divergence structure of the Regge limit can also be
understood via the scattering of massive particles.

\subsection{Regge limit in dimensional regularization}

We recall that in dimensional regularization
the Regge form of $M(s,t)$ 
is \cite{Drummond:2007aua, Naculich:2007ub,DelDuca:2008pj,Naculich:2009cv}
\be
M(s,t) 
=
\beta (t) \left(s\over t\right)^{\alpha(t)-1}
\label{reggetraj}
\ee
where the trajectory function is\footnote{
The tree amplitude supplies another power of $s/t$ so that,
in the Regge limit, the color-ordered amplitude $A \arg $  
goes as $(s/t)^{\alpha(t)}$. }
\ba
\alpha(t)  - 1
&=& 
\suml  a^\ell \left[
{\gamma^\pel \over 4 \ell \eps} 
+ {\cG_0^\pel \over 2} 
\right]
       \left( \mu^2 \over t \right)^{\ell\eps}
	+ \cO(\eps) 
\ea
and the form of the residue 
$ \beta (t) $ may be found in ref.~\cite{Naculich:2007ub,Naculich:2009cv}.

The form of \eqn{reggetraj} suggests that the IR-finite contribution
to the $L$-loop amplitude $M^\Ell(s,t)$
grows as $\log^L  (s)$ in the Regge limit.
This conclusion, however, is not quite warranted
because the neglected $\cO(\eps)$ and higher terms 
could in principle conspire with the pole terms to yield higher powers
of $\log (s)$.  
Nevertheless, it was shown 
that this does not occur
at least through three loops \cite{Naculich:2009cv}, 
and that the $L$-loop amplitude in fact grows as $\log^L(s)$.
It is not unreasonable to hope that this should hold to all orders.

On the other hand, it is definitely not true that
the individual $L$-loop diagrams contributing to $M^\Ell(s,t)$
are bounded by $\log^L(s)$ in the Regge limit.   
As an explicit counterexample, consider the 
Regge limit of the contributions to the three-loop amplitude,
keeping only terms of order $\log^3 (s/t)$ or higher: 
\ba
I_{3a}(s,t) &=&  \cO(\logst),
\\
I_{3a}(t,s) &=& 
\left(\mu^2 \over t\right)^{3 \eps}
\left[  
 \frac{1}{\eps^3} \Bigl( - \logst^3 \Bigr)
+\frac{1}{\eps^2} 
         \left( \frac{1}{12} \logst^4\right)
+\frac{1}{\eps} 
	\left( \frac{1}{60} \logst^5 
	+\frac{25\pi^2 }{36} \logst^3 \right)
+\cO(\eps^0)
\right] + \cO(\logst^2),
\nonumber\\
I_{3b}(s,t) &=& 
\left(\mu^2 \over t\right)^{3 \eps}
\left[  
 \frac{1}{\eps^3} \left( - \frac{1}{3} \logst^3 \right)
+\frac{1}{\eps^2} 
         \left( -\frac{1}{12} \logst^4\right)
+\frac{1}{\eps} 
	\left( -\frac{1}{60} \logst^5 
	-\frac{13\pi^2 }{36} \logst^3 \right)
+\cO(\eps^0)
\right] + \cO(\logst^2),
\nonumber\\
I_{3b}(t,s) &=& 
\left(\mu^2 \over t\right)^{3 \eps}
\left[  
 \frac{1}{\eps^3} \left( \frac{1}{6} \logst^3 \right)
+\frac{1}{\eps^2} 
         \left( \frac{1}{24} \logst^4\right)
+\frac{1}{\eps} 
	\left( \frac{1}{120} \logst^5 
	+\frac{13\pi^2 }{72} \logst^3 \right)
+\cO(\eps^0)
\right] + \cO(\logst^2)
\nonumber\ea
where we have denoted $\logst \equiv \log(s/t)$.
Combining these contributions using \eqn{Mhiggsthree},
we obtain the Regge limit of the three-loop amplitude
\be
M^\Three (s,t) = 
\left(\mu^2 \over t\right)^{3 \eps}
\left[
 \frac{1}{6\eps^3} 
-\frac{\pi^2}{24\eps}  
+\cO(\eps^0)
\right]
\logst^3   + \cO(\logst^2)
\ee
but it is clear that the Regge behavior comes from no 
single diagram, and in fact a cancellation of higher
powers of $\log(s/t)$ occurs among the various contributing
diagrams.
This means that in dimensional regularization, it is very difficult to
ascertain the Regge behavior of $M^\Ell(s,t)$ by considering the
Regge behavior of individual diagrams. 

In the next section, we will compare this to the Regge limit in
Higgs regularization. 
As was noted in the introduction, 
there is a subtlety, namely that the Regge limit of individual diagrams
can depend on the order in which the various limits are taken.
As we will see in the next section,
with one way of taking the limits,
the leading Regge behavior of the $L$-loop amplitude is determined
by a single contributing diagram, the vertical ladder diagram.

\subsection{Regge limit in Higgs regularization}

The conjectured exponentiation \eqn{all-loop-higgs} 
of the four-point amplitude in Higgs regularization can be rewritten as
\be
\label{higgsregge}
\log M  (s,t)
=
-  
{1 \over 4} \gamma(a) 
\log \Bigl( \toverm \Bigr) \log\Bigl(\soverm\Bigr)
- \tilde{\cG}_{0} (a) 
\left[ \log \Bigl( \toverm \Bigr)+ \log\Bigl(\soverm\Bigr) \right]
+ 
{\pi^2 \over 8} \gamma(a)
+ \tilde{c}(a)  + \cO(m^2)
\ee
As pointed out in sec.~\ref{sect-exp},
in Higgs regularization
we may simply exponentiate \eqn{higgsregge} to 
obtain the four-point amplitude.
Moreover, \eqn{higgsregge} implies that the amplitude
is Regge exact,\footnote{\Eqn{exact} could also be rewritten with $s$ and
$t$ exchanged, due to the $s \leftrightarrow t$ symmetry of
\eqn{higgsregge}.} up to terms that vanish as $m^2 \to 0$:
\be
\label{exact}
M(s,t) 
= \beta (t) \left( \soverm\right)^{\alpha(t)-1} + \cO(m^2)
\ee
with
\ba
\label{alpha}
\alpha(t) -1
&=&
- {1 \over 4} \gamma(a) \log \Bigl(\toverm\Bigr)
 - \tilde{\cG}_{0} (a) \,,
\\
\label{beta}
\beta(t) &=& 
\exp
\left[ -\tilde{\cG}_{0} (a) \log \Bigl( \toverm\Bigr) 
+ {\pi^2 \over 8} \gamma(a) + \tilde{c}(a)  \right] \,.
\label{higgsregge2}
\ea
The 
$ - {1 \over 4} \gamma (a) \log (t/m^2)$
piece of the trajectory $\al(t)$ agrees with the conjecture
in eq. (2.18) of ref.~\cite{Schnitzer:2007rn},
but in addition there is a contribution 
from the analog of the collinear anomalous dimension.

\Eqn{higgsregge}  implies that the
leading-log (LL) behavior of the $L$-loop amplitude is given by
\be
\label{leadinglog}
M^\Ell (s,t) \regge   {(-1)^L \over L!} 
\log^L  \Bigl( \toverm \Bigr)
\log^L \Bigl(\soverm\Bigr) .
\ee
In fact we can use \eqn{higgsregge} to organize 
the $L$-loop amplitude
in a leading-log expansion in $\log(s/m^2)$,
with LL, NLL, and NNLL terms given by
(recalling that $u\equiv m^2/s$ and $v\equiv m^2/t$)
\ba
\label{ll-expansion}
M^\Ell
&=& 
\left[
{1 \over L!} (-\log v)^L
\right]  \log^L(u)
\label{Lloop}
\nonumber\\
&+&
\left[
	\left( 
		\frac{\pi^2}{2(L-1)!} 
 		+\frac{\gamma^\Two }{4(L-2)!} 
	\right) (-\log v)^{L-1}  
+                 {\tilde{\cG}_0^\Two \over (L-2)!} 
         (-\log v)^{L-2}
\right]  \log^{L-1} (u)
\nonumber\\
&+& 
\Bigg[
       -{\tilde{\cG}_0^\Two \over (L-2)!} 
         (-\log v)^{L-1}
\nonumber\\
&+&
	\left( 
 		\frac{8\tc^\Two +\pi^4 +\pi^2 \gamma^\Two }{8(L-2)!} 
 		+\frac{2\gamma^\Three+\pi^2 \gamma^\Two  }{8(L-3)!} 
 		+\frac{(\gamma^\Two)^2 }{32(L-4)!} 
	\right) 
        (-\log v)^{L-2}  
\nonumber\\
&+&
	\left( 
 		\frac{2\tilde{\cG}_0^\Three +\pi^2\tilde{\cG}_0^\Two }{2(L-3)!} 
 		+\frac{\gamma^\Two \tilde{\cG}_0^\Two }{4(L-4)!} 
	\right) 
         (-\log v)^{L-3}  
+                  { ( \tilde{\cG}_0^\Two )^2  \over {2 (L-4)!} } 
         (-\log v)^{L-4}
\Bigg]  \log^{L-2} (u)
\nonumber\\
&+&\cO( \log^{L-3} (u) )
\ea
where we have explicitly set
$\gamma^\One = 4$ and $\tilde{\cG}_{0}^\One = \tc^\One = 0$.
In fact we have already tested the validity of this expansion 
through three loops in sec.~\ref{sect-exp} by evaluating 
the integrals that contribute to the amplitudes 
in a leading log expansion.

As discussed in the introduction, however,
there is a subtlety in how one evaluates the leading log expansion  
of the integrals contributing to the four-point amplitude.
One approach, which we term (a), is to first evaluate all the integrals 
in the small $m^2$ limit, and subsequently to take the limit $s \gg t$.  
This is what has been done for all integrals up to this point in the paper.  
In this section, we will explore another approach, which we call (b), 
in which we evaluate the $s \gg t$ limit of the integrals first,
for finite $m^2$, and subsequently take the small $m^2$ limit of the
coefficients of the leading log expansion.
We will find that, although the intermediate details differ,
the final result is still given (at least through three loops)
by \eqn{ll-expansion}.

For one and two loops, we have verified that both methods (a) and (b) yield
the same results for the integrals
namely, eqs.~(\ref{I1regge}),
(\ref{I2reggeu}), and (\ref{I2reggev}).
At three loops, however, the results begin to differ.
The three-loop horizontal ladder diagram (cf. fig.~\ref{fig-23loop}),
gives\footnote{The coefficients of the
$\pi^4$ terms were obtained numerically,
and replaced by their probable rational equivalents.}
\begin{eqnarray}
\label{I3areggeuB}
&&\hspace{-0.5in}\Reggeb
 I_{3a}(u,v) 
=
\log(u) \bigg[ \frac{4}{15} \log^5(v) + \frac{8}{9} \pi^2 \log^3(v) + \frac{28}{45} \pi^4 \log(v) + \cO(v) \bigg]  \\
&+& \left[ -\frac{7}{90} \log^6(v) - \frac{7}{9} \pi^2 \log^4(v) - \frac{8}{3} \zeta_{3} \log^3(v)  +  
\frac{58}{45} \pi^4 \log^2(v) - 35.786 \log(v) -323.7  +\cO(v) \right] \nonumber \\
&+& \cO(u) \,, \nonumber
\end{eqnarray}
which in fact agrees with \eqn{I3areggeu}.
The vertical three-loop ladder, however, gives in the (b) limit
\begin{eqnarray}
&&\hspace{-0.5in}\Reggeb
 I_{3a}(v,u) =
\log^3(u) \, \left[ \frac{4}{3} \log^3 (v)  + \cO(v) \right]  
\\ &+& 
\log^2(u) \, \left[ -\frac{8}{3} \log^4(v) - \frac{10}{3} \pi^2 \log^2(v) - 8 \zeta_{3} \log(v) +\cO(v) \right] \nonumber \\
 &+& \log(u) \, \left[ \frac{34}{15} \log^5(v) + \frac{64}{9} \pi^2 \log^3(v) + 8 \zeta_{3} \log^2(v) + \frac{34}{15} \pi^4 \log(v) + 75.717 +\cO(v) \right]  \nonumber \\
 &+& \left[ -\frac{34}{45} \log^6(v) - \frac{35}{9} \pi^2 \log^4(v)  - \frac{8}{3} \zeta_{3} \log^3(v) - \frac{176}{45} \pi^4 \log^2(v) - 111.504 \log(v) -363.4 +\cO(v) \right] \nonumber \\
&+& \cO(u) \,.
\nonumber
\end{eqnarray} 
In the other way (a) of taking the limit, the leading term goes as $\log^6(u)$.
For the tennis court diagram in the orientation shown in  fig.~\ref{fig-23loop}
we obtain
\begin{eqnarray}
&&\hspace{-0.5in}\Reggeb
 I_{3b}(u,v) =
\log^2(u) \, \left[ \frac{4}{3} \log^4 (v)  + \frac{4}{3} \pi^2 \log^2(v) + \cO(v) \right]  \\
 &+& \log(u) \, \left[ -\frac{9}{5} \log^5(v) - \frac{44}{9} \pi^2 \log^3(v) - 8 \zeta_{3} \log^2(v) - \frac{76}{45} \pi^4 \log(v)  -24.886 +\cO(v) \right] \nonumber \\ 
&+&  \left[ \frac{43}{60} \log^6(v) + \frac{32}{9} \pi^2 \log^4(v) +\frac{16}{3} \zeta_{3} \log^3(v)  - \frac{143}{45} \pi^4 \log^2(v) + 111.499 \log(v) +298.5 +\cO(v) \right]  \nonumber \\
 &+&\cO(u) \,,
\nonumber
\end{eqnarray}
while for the tennis court in the transposed orientation, we obtain
\begin{eqnarray}
\label{I3breggevB}
&& \hspace{-0.5in}\Reggeb
I_{3b}(v,u) =
 \log(u) \, \left[ \frac{8}{15} \log^5(v) + \frac{8}{9} \pi^2 \log^3(v) + \frac{16}{45}\pi^4 \log(v) + \cO(v) \right]  \\
 &+& \left[ -\frac{3}{10} \log^6(v) -\frac{11}{9}\pi^2 \log^4(v) -\frac{8}{3} \zeta_{3} \log^3(v) -\frac{46}{45} \pi^4 \log^2(v) -24.888 \log(v) + 28.5  +\cO(v) \right] \nonumber \\
&+&\cO(u) \,. \nonumber
\end{eqnarray}
Thus, the form of the leading log expansion of the three-loop integrals 
depends on the order in which the limits are taken.

Despite the different expressions for the individual integrals,
we find that the three-loop amplitude itself is 
the same in both Regge (a) and (b) limits.
Plugging the expressions above into \eqn{Mhiggsthree}, 
we find that the expression (\ref{fullMthree}) is reproduced,
using the coefficients in \eqns{twoloopanom}{threeloopanom}, 
just as when we evaluated the integrals in the other order of limits.
In sec.~\ref{sec-bhabha}, we give an heuristic argument
which helps to explain why the Regge limit of the amplitude should be
independent of the order in which the limits are taken.

A significant difference between these two approaches is that
while, in the Regge (a) limit, most of the diagrams contribute
to  the leading log term (\ref{leadinglog}) of the amplitude
(as is also the case in dimensional regularization),
in the Regge (b) limit,
only one diagram, the vertical ladder,
contributes to the leading-log behavior 
of the amplitude in the Regge limit.
Using
\be
M^\Ell = \left( - {1 \over 2} \right)^L 
\Bigl[ I_{La}(v,u) + {\rm ~~(all~other~} L{\rm-loop~diagrams)~~} \Bigl] 
\ee
one can show that the leading log behavior of the $L$-loop
amplitude (\ref{leadinglog}) 
can be wholly accounted for 
by the vertical $L$-loop ladder diagram $I_{La}(v,u)$,
suggesting that all other diagrams are subdominant 
(\ie, grow no faster than $\log^{L-1} (s)$)
in the Regge (b) limit.
Using the methods of ref.~\cite{Eden}, we derive in appendix \ref{sect-eden}
the LL, NLL, and NNLL contributions of the vertical $L$-loop diagram,
obtaining 
\ba
\label{nnll}
&&\hspace{-0.5in}\Reggeb
I_{La} (v,u)
= 
  {2^L \over L!} \log^L u \log^L v 
\nonumber
\\
&-&
 {2^L \over (L-1)!} \log^{L-1} u 
		 \left[ {1\over 3}  (L-1)  \log^{L+1} v
       			+  (L+2) \zeta_2   \log^{L-1} v
       			+  (L-1) \zeta_3 \log^{L-2} v \right]
\nonumber\\
&+&
 {2^L \over (L-2)!} \log^{L-2} u 
 	\bigg[ {10 L^2 - 14 L + 3 \over 180} \log^{L+2} v
	+	{(L+1)^2 \over 18} \pi^2 \log^{L} v
\nonumber\\
&+&
		{ L(L-2) \over 3} \zeta_3 \log^{L-1} v
	+	{(L+3)(5L+2) \over 360} \pi^4 \log^{L-2} v
	+ {L-2 \over 6} (L + 1.826) \pi^2 \zeta_3  \log^{L-3} v
\nonumber\\
&+&
		{(L-2)(L-3)\over 2} \zeta_3^2 \log^{L-4} v
\bigg]
+ \cO( \log^{L-3} u ) \,.
\ea
It is clear from comparing \eqns{ll-expansion}{nnll}
that other diagrams begin to contribute at the NLL
order.
It is difficult, however,
to use the methods of ref.~\cite{Eden}
to estimate the leading logarithmic growth for the other diagrams 
due to the presence of numerator factors in the integrals.
(It would be interesting to extend the methods of ref.~\cite{Eden}
to these cases.)
We suspect that only a small subset of the allowed $L$-loop diagrams
contribute to the NLL terms of $M^\Ell$;
it would be nice to characterize which.

\subsection{Regge limit in cutoff regularization}

Mandelstam \cite{Mandelstam:1965zz} has given certain criteria for
establishing whether Reggeization occurs, i.e. whether an elementary
field in a Lagrangian field theory lies on a Regge trajectory or not.
It depends on a ``counting procedure,'' which must be carried out
separately for each field of the Lagrangian, where for renormalizable
theories one simply does the counting at $j=0$, 1/2, or 1 to consider
the issue.  It has been shown \cite{Grisaru:1973vw,Grisaru:1974cf,Grisaru:1979wi} that the 
elementary fields of renormalizable Yang-Mills gauge theories 
lie on a Regge trajectory if the theory has a mass gap, 
where theories with a Higgs mechanism
provide a case in point.  These arguments are purely local in nature,
in that the trajectory of a gluon passes through $j=1$ at the mass of
the particle, but no global information about the trajectory is provided
in this construction.

In addition to the ``counting criteria,'' a certain factorization
condition among the helicity matrices of the scattering process must be
satisfied if the Reggeization is to take place.  This requirement results
from the solution of the unitarity-analyticity equations satisfied by
the scattering amplitude analytically continued in angular momentum. As
a result an integral equation must be satisfied, where the potential $V$
is the inhomogeneous term in the integral equation. If $V$ is taken to
be the kinematical-singularity-free, partial-wave projection of the Born
approximation helicity amplitude, the solution to the integral equation
provides an analytic form for the Regge trajectory which is equivalent to
the LL approximation to the trajectory function.\footnote{See sec. $2$
of ref.~\cite{Grisaru:1982bi}.}

Given this background, in ref.~\cite{Schnitzer:2007rn} the Regge limit of the ladder approximation for the color-ordered,
tree-approximation-stripped amplitude was considered, 
with cutoff regularization 
(\ie, all the propagators in the loop integrals are given a mass $m$) 
for gluon-gluon
scattering in $\mathcal{N}=4$ SYM. It was shown that
\begin{equation}\label{how-1}
M(s,t) \regge 
\beta(t) \left( \frac{s}{m^2} \right)^{\alpha(t)-1}\,,
\end{equation}
where
\begin{equation} \label{how-2}
\alpha(t) = 1 - a \log \Bigl(  \toverm \Bigr) + \cO(a^2) \,.
\end{equation}
Hence \eqn{how-2} agrees with \eqn{higgsregge2}
to lowest order in the coupling.
The reason for this is that, as we saw above, 
(1) the leading log contribution to the $L$-loop amplitude in the 
Regge (b) limit in Higgs regularization comes from the 
vertical ladder diagram alone, and
(2) as we show in appendix~\ref{sect-eden},
the LL contribution of the vertical ladder
is independent of the masses of the propagators of the rungs of the ladder,
which is the only difference between 
the Higgs- and cutoff-regulated diagrams.

In ref.~\cite{Schnitzer:2007rn} it was conjectured that with the cutoff regularization
\begin{equation}\label{how-3}
\alpha(t) = 1 - \frac{1}{4} \gamma(a) \, \log \Bigl( \toverm \Bigr) \,,
\end{equation}
which satisfies 
\begin{equation}\label{how-4}
\alpha(m^2) =1 \,,
\end{equation}
as required by general principles. 
Note that \eqn{how-3} coincides with \eqn{higgsregge2} 
from the Higgs regularization up to a scheme-dependent constant,
which begins at ${\cal O}(a^2)$.

While Higgs and cutoff regulators yield
identical results at the leading log level,
one could ask whether they differ at the subleading log level.
Consider the diagrams that contribute to the two-loop 
amplitude (\ref{Mhiggstwo}).
Whereas in Higgs-regularization the middle line is massless,
in the cutoff scheme,
all internal propagators have a uniform mass $m$.\footnote{Note 
that, in contrast with dimensional or Higgs regularization,
it is not clear whether other integrals might not also contribute 
in cutoff-regularization, as could happen when one goes off-shell.} 
To compute the two-loop diagram with uniformly massive propagators, 
more MB integrals are needed relative to the Higgs-regulated case. 
For example, a naive use of 
AMBRE \cite{Gluza:2007rt}
yields a nine-fold MB representation. 
Using the methods described in appendix \ref{sect-MB}, 
we have written down a seven-fold MB representation, 
from which we extract the following results
in the Regge (b) limit $u \ll v \ll 1$:
\begin{eqnarray}
\Reggeb I_{2;\, {\rm cutoff}}(u,v)  &=& 
\log(u) \left[ \frac{4}{3} \log^3(v) + \frac{4}{3} \pi^2 \log(v) + 4 \zeta_{3} + \cO(v) \right] +  \cO(\log^{0}(u))\,,
\nonumber\\
\Reggeb I_{2; \,{\rm cutoff}}(v,u) & = & \log^2(u) \bigg[ 2 \log^2(v) + O(v) \bigg] 
\\ &+& 
\log(u) \left[ -\frac{4}{3} \log^3(v) -\frac{8}{3} \pi^2  \log(v) 
+ \frac{4}{3} \zeta_{3}  + \cO(v) \right] 
+ \cO( \log^{0}(u)) \,.
\nonumber
\end{eqnarray}
Comparing this with \eqns{I2reggeu}{I2reggev},
we see that the $\zeta_3$ coefficients in the $\log(u)$ terms 
are changed.
There are presumably also changes in the uncomputed  $\log^{0}(u)$ terms.

We can also compute the NLL terms of 
the cutoff-regulated $L$-loop vertical ladder diagram.   The
calculation of appendix \ref{sect-eden} is unchanged,
except that the middle rungs of the $(2+2\eps)$-dimensional
diagrams depicted in fig.~\ref{fig-regge2} become massive.
As a result, the integral $B_2$ in \eqn{BT}, for example, is changed to 
\be
e^{-2 \epsilon \gamma_{\rm E} } \, B_{2;\,{\rm cutoff}} 
=  4 \log^2 u 
+ \epsilon \left[ \frac{10}{3} \log^3 u + \frac{4}{3} \pi^2 \log u 
- \frac{4}{3}  \zeta_{3} \right]  + \ldots
\ee
as a consequence of which we have
\ba
&&\hspace{-0.5in}\Reggeb
I_{La;\,{\rm cutoff}} (v,u) = 
  {2^L \over L!} \log^L u \log^L v 
\nonumber\\
&-&
 {2^L \over (L-1)!} \log^{L-1} u 
\left[ {1\over 3}  (L-1)  \log^{L+1} v
+  (L+2) \zeta_2   \log^{L-1} v
- \frac{1}{3} (L-1) \zeta_3 \log^{L-2} v \right]
\nonumber\\
&+& \cO( \log^{L-2} u ) 
\ea
which should be compared with \eqn{nnll}.

\subsection{Scattering of massive particles and Regge behavior}
\label{sec-bhabha}

In this section, we present a different approach 
which helps to explain why we obtained the same 
three-loop amplitude from \eqn{Mhiggsthree} 
in both Regge (a) and (b) limits, despite the fact that
the contributing integrals 
(\ref{I3areggeu}-\ref{I3breggev})
and (\ref{I3areggeuB}-\ref{I3breggevB})
differ so drastically from one another.

Consider the four-point amplitude on the Coulomb branch
of $\cN=4$ SYM theory, as reviewed in sec.~\ref{sect-revdcs}, 
with the scattered particles satisfying the on-shell conditions
\begin{equation}
 p_{i}^2 = - (m_{i} - m_{i+1})^2\,.
\end{equation}
Recall that the assumption of dual conformal symmetry 
implies that the amplitude is a function of two variables
$u$ and $v$, defined in \eqn{defuv}.
In previous sections of this paper, 
we considered the equal mass case, $m_{i} = m$, so that 
$u=m^2/s$ and $v=m^2/t$, and the external states are massless
$p_i^2 = 0$.
The Regge limit $s \gg t \gg m^2$ corresponds to $u \ll v \ll 1$.

\begin{figure}
\psfrag{A}[cc][cc]{$(a)$}
\psfrag{B}[cc][cc]{$(b)$}
\psfrag{p1}[cc][cc]{$p_1$}
\psfrag{p2}[cc][cc]{$p_2$}
\psfrag{p3}[cc][cc]{$p_3$}
\psfrag{p4}[cc][cc]{$p_4$}
 \centerline{
 {\epsfxsize13cm  \epsfbox{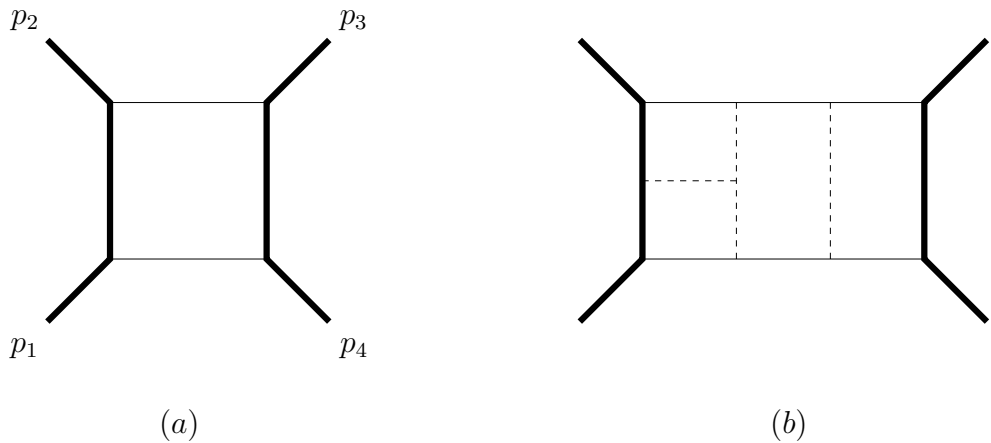}}
}
\caption{\small
(a) One-loop scattering of massive particles. 
(b) Sample higher-loop diagram. Fat lines and thin lines denote particles
of mass $M$ and $m$, respectively. 
Dashed thin lines denote massless particles.}
\label{fig-bhabha}
\end{figure}

In this section, we consider instead the two mass case, 
$m_{1}=m_{3}=m$ and $m_{2}=m_{4}=M$, 
which implies $p_{i}^2= -(M-m)^2$,
and (cf. \eqn{defuv})
\be
u = \frac{m^2}{s}
\qquad{\rm and} \qquad v = \frac{M^2}{t}. 
\ee
Dual conformal symmetry implies that the Regge 
limit $u \ll v \ll 1$ can be attained
by choosing $m^2 \ll M^2 \ll s,t$,
so that
the on-shell condition becomes $p_{i}^2 \approx -M^2$.
This corresponds to the scattering of particles of mass $M$ 
by the exchange of particles of much lighter mass $m$.
See fig.~\ref{fig-bhabha}(a) for the one-loop contribution
to this scattering. 
At higher loops, massless particles will also be 
exchanged in the interior of the diagram; 
see fig.~\ref{fig-bhabha}(b) for a sample higher-loop diagram. 

As was discussed in the introduction and in section \ref{sect-regge}
it is important to specify in which order the Regge limit is taken.
In this section, we will first take $m^2$ small for fixed $M^2$, $s$, and $t$.
This corresponds to the first part of Regge limit (b), 
namely, $u \to 0$, i.e., $u \ll v$ and $u \ll 1$.
Later in this section, we will take $M^2 \ll s$, $t$,
which corresponds to the second part ($v \ll 1$).
The amplitudes discussed above have soft IR divergences (as $m \to 0$), 
but no collinear divergences, 
so the double logarithms $\log^2(m^2)$ characteristic of overlapping 
divergences are absent, leaving only single logarithms $\log(m^2)$. 
At one loop one can show that 
\begin{equation}
M^{(1)}(u,v) 
\quad\mathrel{\mathop{\longrightarrow}\limits_{u \to 0 }}\quad
-\frac{1}{2} (\log u)  \, \Gamma_{\rm cusp}(a, \theta)\big|_{ {\rm 1-loop}} + \cO(u^0)
\end{equation}
where $\Gamma _{\rm cusp}(a, \theta)$ 
is the cusp anomalous 
dimension \cite{Polyakov:1980ca,Korchemsky:1985xj,Ivanov:1985np,Korchemsky:1987wg}
of a Wilson line with a non-light-like cusp, and
with $\theta$ being the (Minkowskian) cusp angle defined by 
$\cosh \theta = p_{2} \cdot p_{3} / M^2 $.
It is plausible that the leading behavior for all loops is
\begin{equation}\label{all-loop-Bhabha}
M(u,v) 
\quad\mathrel{\mathop{\longrightarrow}\limits_{u \to 0 }}\quad
 \exp \left\{ -\frac{1}{2} (\log u) \, \Gamma_{\rm cusp}(a, \theta) + \cO(u^0) \right\} \,.
\end{equation}

Next, we take the further limit $M^2 \ll t$, \ie, $v \ll 1$.
In this limit, the cusp anomalous dimension goes 
to \cite{Korchemsky:1991zp}
\begin{equation}\label{lim-gammacusp}
\Gamma_{\rm cusp}(a, \theta) 
\quad\mathrel{\mathop{\longrightarrow}\limits_{v \to 0 }}\quad
(\log v) \, \Gamma_{\rm cusp}(a)  + \cO(v^0)\,,
\end{equation}
where $\theta \approx - \log v$ and $\Gamma_{\rm cusp}(a) = \gamma(a)/2$. 
Accepting the conjectured \eqn{all-loop-Bhabha}, 
one obtains in the Regge limit
\begin{equation}\label{all-loop-Regge}
\Reggeb M(u,v) 
= \exp \left\{ -\frac{1}{2} (\log u)(\log v) 
\Gamma_{\rm cusp}(a) + \cdots \right\} \,.
\end{equation}
Since we have taken the limits in the order $u \ll 1$ with $v$ fixed, 
followed by $v \ll 1$, this is Regge limit (b)
as defined in the introduction.
Note that we arrive in this way at the 
same (leading log) result as \eqn{higgsregge},
which was obtained in Regge limit (a), \ie,
taking $u$, $v\ll 1$ first, followed by $u \ll v$.
This helps to explain why we obtained the 
same three-loop amplitude from \eqn{Mhiggsthree} 
in the two Regge limits,
despite the fact that the contributing integrals
had different forms in these two limits. 

It is also noteworthy that \eqn{all-loop-Regge} directly 
exhibits the single log behavior expected of the Regge limit, 
without the cancellation of double logs that occurs 
in taking the Regge limit (\ref{higgsregge}) 
of the exponential ansatz (\ref{all-loop-higgs}).

\section{Discussion}
\setcounter{equation}{0}

In this paper we have tested various conjectures and issues related to the
Higgs regulator scheme proposed in ref.~\cite{Alday:2009zm}  for $\cN=4$
SYM theory.  The assumption of exact dual conformal symmetry enabled us
to compute the four-gluon scattering amplitude in the planar limit to
three-loop accuracy.
Our results are consistent,
in various limits,
with an analog of the BDS conjecture 
proposed for the Higgs-regulated four-gluon amplitude,
with the same value (to three loops) 
of the cusp anomalous dimension 
as that found in dimensional regularization. 
This is as expected since the cusp anomalous dimension is 
IR-scheme-independent.

The Regge limit of the four-point function was considered in various
IR regulator schemes.  
It was shown that the leading log behavior of the Regge
trajectory is determined by the sum of vertical ladder graphs for
the Higgs regulator and for a cut-off regulator.\footnote{Here we 
are referring to the Regge (b) limit, as defined in the paper.}
This contrasts with dimensional regularization, 
where no single set of diagrams dominates
at any order of perturbation theory.  
Further one may associate the perturbative expansion 
for the cusp anomalous dimension 
with the NLL, NNLL, $\cdots$ approximations to the gluon Regge trajectory.
A particular dual conformal mass assignment allows one to obtain directly
the single logarithmic behavior of the Regge trajectory, 
without the necessity of cancellation of $\log^2$ terms.

We also found in the course of this work that Higgs-regulated amplitudes lead
to more efficient integral representations (\ie, fewer MB parameters)
than more generic (\eg, cutoff-regulated) schemes.

There are several issues deserving further attention.
It would be useful to extend the methods of ref.~\cite{Eden}
to determine the Regge limits of diagrams involving non-trivial 
(\ie, loop-momentum-dependent) numerator factors.
It would also be  interesting to determine whether 
there is a simple characterization of the subset of $L$-loop
diagrams that contribute to the NLL, NNLL, $\cdots$
terms in the $L$-loop amplitude in the Regge limit.

Overall, we have seen that the Higgs regulator for planar
$\cN=4$ SYM amplitudes has a number of advantages over other
regulators.

\section*{Acknowledgments}

It is a pleasure to thank 
Z.~Bern, L.~Dixon, D.~Freedman, G.~Korchemsky,
S.~Moch, H.~Nastase, J.~Plefka, R.~Roiban, 
C.~Tan, and A.~Volovich 
for stimulating discussions.
J.H. thanks T.~Riemann, S.~Weinzierl, and V.~Yundin  for information
on refs.~\cite{Gluza:2007rt,Czakon:2005rk,Bogner:2007cr},
and the theoretical physics group at Brandeis University, 
where part of this work was done, for hospitality.

\appendix

\section{Mellin-Barnes representations of the integrals}
\setcounter{equation}{0}
\label{sect-MB}

In this section we derive Mellin-Barnes (MB) representations for the
integrals considered in the paper.  In doing so we follow the loop-by-loop
approach advocated in refs.~\cite{Usyukina:1992jd,Smirnov:2006ry}.
In this approach, one successively derives MB representations for one-loop
subintegrals.  Remarkably, the Higgs masses can be incorporated rather
naturally into this procedure.

Tools that we have found useful are the Mathematica packages MB
\cite{Czakon:2005rk} and AMBRE \cite{Gluza:2007rt}. The latter can also
be used to derive MB representations.  
Of course, one can sometimes find
MB representations involving fewer MB integrals
by going through the derivation by hand.  
Since the number of integrals we need is rather small we have
followed the latter strategy. We present details of the derivation below,
since the essential steps when deriving MB representations for higher-loop
or higher-point integrals are the same.

It is interesting to note that using a uniform cut-off (i.e., giving
a mass to {\it all} propagators of the integral) makes the resulting
MB representations considerably more intricate, and one needs more
MB parameters.

\subsection{MB representation for the $3$-loop ladder diagram}
\begin{figure}
%\vspace*{10mm}
%
\psfrag{a}[cc][cc]{$(a)$}
\psfrag{b}[cc][cc]{$(b)$}
\psfrag{x1}[cc][cc]{$x_{1}$}
\psfrag{x2}[cc][cc]{$x_{2}$}
\psfrag{x3}[cc][cc]{$x_{3}$}
\psfrag{x4}[cc][cc]{$x_{4}$}
\psfrag{x5}[cc][cc]{$x_{5}$}
\psfrag{x6}[cc][cc]{$x_{6}$}
\psfrag{x7}[cc][cc]{$x_{7}$}
\centerline{
{ \epsfxsize15cm  \epsfbox{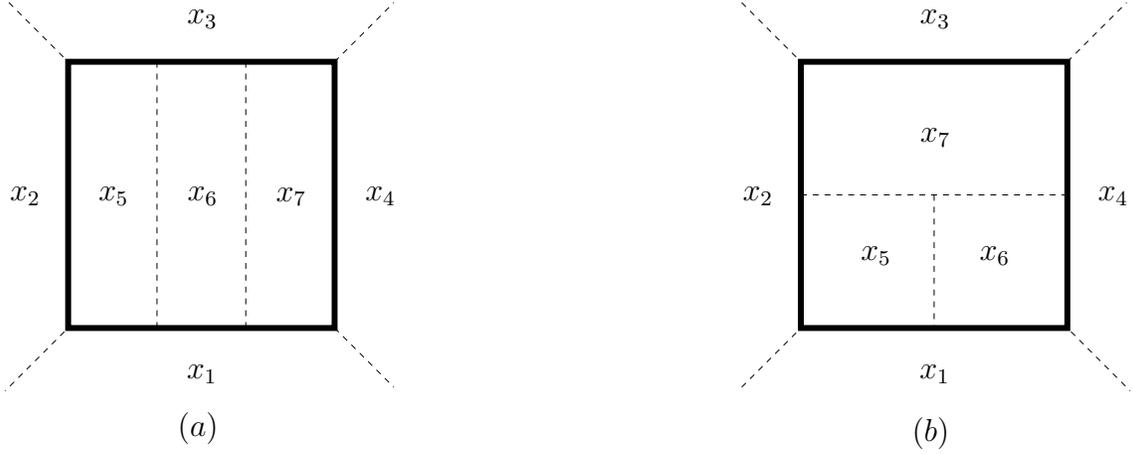}}
} \caption{Diagrams of two dual conformal integrals at three loops. 
Thick lines correspond to massive propagators, thin lines to massless ones. 
The $x_{i}$ are dual coordinates. 
Figure (a) shows the three-loop ladder diagram, 
with a factor of $s^3 t = (x_{13}^2)^3 x_{24}^2$ removed. 
Figure (b) shows the tennis court diagram, 
with a factor of $s t^2$ removed and a numerator $x_{17}^2 + m^2$ 
not shown in the diagram.}
\label{fig-threeloop}
\end{figure}

Let us consider the three-loop ladder depicted in figure \ref{fig-threeloop}a,
\begin{equation}\label{int-tripleladder}
I_{3a} = s^3 \, t\, \int \frac{d^{4}x_{5} d^{4}x_{6} d^{4}x_{7}}{( i \pi^2 )^3} \, \left( P_{25,m}\, P_{15,m}\,P_{35,m}\,P_{56}\,P_{36,m}\,P_{16,m} \,P_{67}\,P_{37,m}\,P_{17,m}  \,P_{47,m} \right)^{-1}\,.
\end{equation}
Here $P_{ij,m} = x_{ij}^2 + m^2$ and $x_{i}$ is the dual notation introduced 
in sec.~\ref{sect-revdcs}.
In the massless case, it is possible to derive a convenient MB representation by successively doing the loop integrations \cite{Usyukina:1992jd,Smirnov:2006ry}.
As we will see presently, the same strategy also works well in the massive case.
We begin with the $x_{5}$ subintegral
\begin{eqnarray}
I_{3a}^{(1)} &=&   \int \frac{d^{4}x_{5}}{i \pi^{2}} \, \left( P_{15,m}\, P_{25,m}\,P_{35,m}\,P_{56} \right)^{-1} 
\end{eqnarray}
where 
\begin{eqnarray}\label{ladder-it1}
I_{3a}  &=&  s^3 \, t \,   \int  \frac{ d^{4}x_{6} d^{4}x_{7} }{(i \pi^2)^2}  \,  I_{3a}^{(1)} \,\left( P_{36,m}\,P_{16,m} \,P_{67}\,P_{37,m}\,P_{17,m}  \,P_{47,m} \right)^{-1}\,.
\end{eqnarray}
Introducing $\alpha$ parameters 
(see for example ref.~\cite{Smirnov:2006ry}), we obtain
\begin{eqnarray}
I_{3a}^{(1)} &=& \int_0^{\infty} \frac{ d\alpha_{i} \, \delta( \sum \alpha_{i} -1 )  }{ \left[ \alpha_1 \alpha_3 s + \alpha_6 \left( \alpha_1 P_{16,m} + \alpha_2 P_{26,m} + \alpha_3 P_{36,m} \right) + (\alpha_1 + \alpha_2 + \alpha_3 )^2 m^2 \right]^{2}}\,.
\end{eqnarray}
Note that the range of the sum in the delta function can be chosen arbitrarily.
Here, it is convenient to choose $\sum_{i=1}^3$ in order to simplify the $m^2$ term.
Performing the $\alpha_{6}$ integration,  we obtain 
\be
I_{3a}^{(1)} 
 = \int_0^{1} \, 
\frac{ d\alpha_{i} \, \delta( \sum_{i=1}^{3} \alpha_{i} -1 ) } 
{    (\alpha_1 P_{16,m} + \alpha_2 P_{26,m} + \alpha_3 P_{36,m}) 
      (\alpha_1 \alpha_3 s+  m^2) } \,.
\ee
To perform the remaining integrations over the $\alpha$ parameters,
we introduce three MB parameters.
Using  \eqns{MB-12}{MB-13},
we obtain
\begin{eqnarray}
I_{3a}^{(1)}  &=&  \int \frac{ dz_{i}}{  (2 \pi i)^{3}}  \Gamma(-z_1 ) \Gamma( -z_2 ) \Gamma(-z_3 ) \Gamma(1+z_1 +z_3 ) \Gamma(1 + z_2 ) \, ( m^2 )^{-1  -z_2 }  \nonumber \\
&&\times \,  \int d\alpha_{i} \, \delta( \sum_{i=1}^{3} \alpha_i -1) \, ( \alpha_1 \alpha_3 s )^{z_2 } (\alpha_2 P_{26,m} )^{-1-z_1 -z_3 }  (\alpha_1 P_{16,m} )^{z_1 } (\alpha_3 P_{36,m} )^{z_3 }   \,. 
\end{eqnarray}
The parameter integrals are now easily done with the help of (\ref{alpha-01}), leading to
\begin{eqnarray}\label{ladder-it1-result}
I_{3a}^{(1)}  &=&   \int \frac{dz_{1,2,3}}{(2 \pi i)^{3} }  f^{(1)}(z_{1}, z_{2},z_{3} ) \, s^{z_2 } (m^2 )^{-1 -z_2 }  (P_{26,m})^{-1-z_1-z_3 }   (P_{16,m})^{z_1 } (P_{36,m})^{z_3 }   \,,
\end{eqnarray}
where
\begin{eqnarray}
f^{(1)}(z_{1}, z_{2},z_{3} ) &=& \Gamma(-z_1 ) \Gamma( -z_2 ) \Gamma(-z_3 ) \Gamma(1+z_1 +z_3 ) \Gamma(1 + z_2 ) \nonumber \\
&&\times \,  \frac{ \Gamma(-z_1 - z_3 ) \Gamma(1+z_2+z_1 ) \Gamma( 1+z_2 + z_3 )}{\Gamma(2 (1+ z_2 ))} \,.
\end{eqnarray}
Plugging (\ref{ladder-it1-result}) into (\ref{ladder-it1}),
we obtain
\be
I_{3a}  =  s^3 \, t \,   \int \frac{dz_{1,2,3}}{ (2 \pi i)^{3}} \,f^{(1)}(z_{1}, z_{2},z_{3} ) \, s^{z_2 } (m^2 )^{-1 -z_2 }  \,  \int  \frac{d^{4}x_{7}}{i \pi^2 }   \,  I_{3a}^{(2)} \,\left( \,P_{37,m}\,P_{17,m}  \,P_{47,m} \right)^{-1}\,,
\ee
where $I_{3a}^{(2)}$ is the integral over $x_6$:
\begin{eqnarray}
I_{3a}^{(2)} &=&   \int \frac{ d^{4}x_{6}}{i \pi^2}  \, (P_{26,m})^{-1-z_1-z_3 }   (P_{16,m})^{-1+z_1 } (P_{36,m})^{-1+z_3 }  (P_{67})^{-1}  \nonumber \\
&=& \int_0^{\infty} \frac{ d\alpha_{i} \, \delta( \sum \alpha_{i} -1 ) \, \alpha_{1}^{-z_{1}} \, \alpha_{3}^{-z_{3}}\, \alpha_{2}^{z_{1}+z_{3}} }{ \left[ \alpha_1 \alpha_3 s + \alpha_7 \left( \alpha_1 P_{17,m} + \alpha_2 P_{27,m} + \alpha_3 P_{37,m} \right) + (\alpha_1 + \alpha_2 + \alpha_3 )^2 m^2 \right]^{2}} \nonumber \\
&& \times \frac{1}{\Gamma(1-z_{1}) \Gamma(1-z_{3}) \Gamma(1+z_{1} +z_{3})}
\end{eqnarray}
This is a generalization of the integral considered before with more general powers of the propagators. 
The calculation is completely analogous, and we obtain
\be
I_{3a}^{(2)} =  \, \int \frac{dy_{1,2,3}}{(2\pi i)^{3}} \, f^{(2)}(z_{1,2,3};y_{1,2,3})  \, s^{y_{2}}\, (m^2)^{-1-y_{2}}  \, (P_{17,m})^{y_{1}}  \, (P_{37,m})^{y_{3}}\,  (P_{27,m})^{-1-y_1 - y_3 } \,,
\ee
with
\begin{eqnarray}
f^{(2)}(z_{1,2,3};y_{1,2,3}) &=&   \frac{ \Gamma(-y_1 ) \Gamma(-y_2 ) \Gamma( -y_3 ) \Gamma(1+y_1 + y_3 ) \Gamma(1 + y_2 )}{\Gamma(1-z_{1}) \Gamma(1-z_{3}) \Gamma(1+z_{1}+z_{3}) } \nonumber \\
&& \times \frac{ \Gamma(1-z_1 + y_1 + y_2 ) \Gamma( z_1 + z_3 - y_1 - y_3 ) \Gamma(1-z_3 +y_2 + y_3 )}{\Gamma( 2 (1+y_2 ))}\,.
\end{eqnarray}
At this point we note that, for a ladder diagram with more rungs, the next subintegral would 
be of the same type as the preceding one (up to a change of labels of MB parameters), see figure \ref{fig-ladderit}. In the next section, we will use this iterative structure to
write down a MB parametrization for the $L$-rung ladder with $L$ arbitrary.

Coming back to the three-loop ladder, we are left with the final integration over $x_{7}$, namely
\begin{eqnarray}
I_{3a}  &=&  s^3 \, t \,   \int \frac{dz_{1,2,3}}{ (2 \pi i)^{3}} \, \frac{dy_{1,2,3}}{ (2 \pi i)^{3}} \,f^{(1)}(z_{1,2,3}) \,f^{(2)}(z_{1,2,3};y_{1,2,3}) \, s^{z_2 + y_2 } (m^2 )^{-2 -z_2 - y_2 }  \,   I_{3a}^{(3)} \,,
\end{eqnarray}
where
\begin{equation}\label{Iauxdef}
I_{3a}^{(3)} =  \int \frac{ d^{4}x_{7}}{i \pi^2} \,  (P_{27,m})^{-1-y_1 - y_3} \, (P_{17,m})^{-1+y_1 } \, (P_{37,m})^{-1+y_3 } \, (P_{47,m})^{-1} \,. \\
\end{equation}
Introducing parameter integrals we have
\be
I_{3a}^{(3)} = \frac{1}{\Gamma(1-y_{1}) \Gamma(1+y_1 +y_3 ) \Gamma(1-y_3 ) }\,  \int_{0}^{\infty}    \frac{ d\beta_{i} \, \delta(\sum \beta_i -1 ) \, \beta_{1}^{-y_1 } \, \beta_{2}^{y_1 +y_3 } \, \beta_{3}^{-y_3 }}
{\left[\beta_1 \beta_3 s + \beta_2 \beta_4 t + ( \beta_1 + \beta_2 + \beta_3 + \beta_4 )^2 m^2 \right]^{2} } \,.
\ee
Obviously, here it is convenient  to choose $\sum = \sum_{i=1}^{4}$.
Introducing two more MB parameters, we find
\begin{eqnarray}\label{Iauxfinal}
I_{3a}^{(3)}  &=&   \int \frac{dz_{4,5}}{(2 \pi i)^2}\, f^{(3)}(y_{1,2,3};z_{4,5}) \, s^{z_{4}}\, t^{z_{5}}\, (m^2)^{-2 -z_{4}-z_{5}}\,,
\end{eqnarray}
where
\begin{eqnarray}
 f^{(3)}(y_{1,2,3};z_{4,5}) &=& \frac{\Gamma(-z_4 ) \Gamma(-z_5 ) \Gamma(2 + z_4 + z_5 ) }{\Gamma(1-y_1 ) \Gamma(1-y_3 ) \Gamma(1+y_1 +y_3 )  } \nonumber \\
&\times &  \frac{\Gamma(1+z_{5}) \Gamma(1+y_1 +y_3+z_5 ) \Gamma(1-y_1 +z_4 ) \Gamma(1- y_3 +z_4 ) }{\Gamma( 2 (2+z_4 + z_5 ))}\,.
\end{eqnarray}
Putting everything together we arrive at the final result (relabelling $z_{4,5} \to z_{7,8}$ and $y_{1,2,3} \to z_{4,5,6}$)
\begin{equation}\label{MB-tripleladder}
I_{3a}  = 
\int \frac{ dz_{1,2,3,4,5,6,7,8} }{(2 \pi i)^{8}  }
\, f^{(1)}(z_{1,2,3}) \,f^{(2)}(z_{1,2,3};z_{4,5,6})  \, f^{(3)}(z_{4,5,6};z_{7,8}) 
 \left(\frac{m^2}{s} \right)^{-3-z_2 -z_5 -z_7  } \left(\frac{m^2}{t}\right)^{-1-z_8 }\,.
\end{equation}
In  \eqn{MB-tripleladder}  the integration contours are chosen parallel
to the imaginary axis, with the real parts of the integration variables
defined such that all arguments of the $\Gamma$ functions have positive
real part.
An allowed choice is
\begin{equation}
z_1 = -\frac{1}{2}\,, z_2 = -\frac{1}{4}\,, z_3 = -\frac{1}{4}\,, 
z_4 = -\frac{3}{4}\,, z_5 = -\frac{1}{16} \,, z_6 = -\frac{15}{32} \,, z_7 = -\frac{9}{32}\,, 
z_8 = -\frac{13}{32}\,.
\end{equation}

\subsection{MB representation for the $L$-loop ladder diagram}
\begin{figure}
%\vspace*{10mm}
%
\psfrag{p1}[cc][cc]{$1$}
\psfrag{power1}[cc][cc]{$1+z_1+z_3$}
\psfrag{power2}[cc][cc]{$1-z_1$}
\psfrag{power3}[cc][cc]{$1-z_3$}
\psfrag{powery1}[cc][cc]{$1+y_1+y_3$}
\psfrag{powery2}[cc][cc]{$1-y_1$}
\psfrag{powery3}[cc][cc]{$1-y_3$}
\psfrag{dors}[cc][cc]{$\ldots$}
\psfrag{equation}[cc][cc]{$\propto \, \int \frac{dy_{1,2,3}}{(2 \pi i)^3}\, f^{(2)}(z_{i};y_{i}) \, (\frac{s}{m^2})^{y_{2}} \, \times $}
\centerline{
{ \epsfxsize18cm  \epsfbox{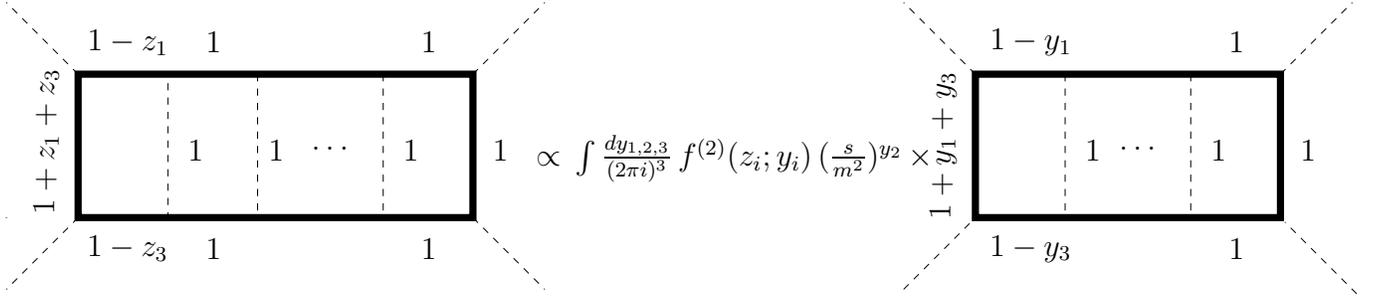}}
} \caption{Diagrammatic representation of the iteration used to derive the $L$-loop MB representation for the ladder integral.
Fat lines correspond to massive propagators, thin lines to massless ones. The numbers indicate the powers of the propagators.
One can see that after removing a rung of the ladder (by introducing three Mellin-Barnes integrals) we obtain an integral of the same
type, and the procedure can be iterated.}
\label{fig-ladderit}
\end{figure}
As was noted in the previous section, one can straightforwardly write down an iterated MB formula for the $L$-loop ladder diagram.
The decisive step is illustrated in figure \ref{fig-ladderit}.
Let us define 
\be
\label{defladders}
I_{La}  =  s^L \, t \, \int \prod_{j=5}^{L+4} \left(  \frac{d^{4}x_{j}}{i \pi^2} \right)   \, ( P_{25,m} P_{15,m} P_{35,m} )^{-1}
\prod_{j=5}^{L+3}  ( P_{j,j+1} P_{1,j+1,m} P_{3,j+1,m}   )^{-1}  \, (P_{4,L+4,m} )^{-1}  \,.
\ee
The MB formula we obtain is $(3L-1)$-fold and is given by (for $L>1$)
\begin{eqnarray}
I_{La} &=& \int  \prod_{j=1}^{L-1} \left( \frac{dz^{(j)}_{1,2,3}}{(2 \pi i)^3}  \right)  \frac{dz_{4,5} }{(2 \pi i)^{2}} \,
\left( \frac{m^2}{s} \right)^{-L-\sum_{j=1}^{L-1} z_{2}^{(j)} -z_4 }  \, \left(\frac{m^2}{t}\right)^{-1-z_5 } \nonumber \\
&& \times    f^{(1)}(z^{(1)}_{1,2,3})  \, \prod_{j=1}^{L-2} f^{(2)}(z^{(j)}_{1,2,3} ; z^{(j+1)}_{1,2,3}) \,  f^{(3)}(z^{(L-1)}_{1,2,3} ; z_{4,5})   \,.
\end{eqnarray}
For completeness, we also give a formula for $L=1$, 
\begin{equation}
I_{1a} =I_{1}  = \int \frac{ dz_{1,2}  }{(2 \pi i)^{2}} \, u^{-1 - z_1}  v^{-1 - z_2 }   \frac{\Gamma(-z_1 ) \Gamma^2(1 + z_1 )\Gamma(-z_2 ) \Gamma^2 (1 + z_2 ) \Gamma(
    2 + z_1 + z_2 )}{\Gamma(4 + 2 z_1 + 2 z_2 ) }
\end{equation}
which is equivalent but more symmetric than  
the one given in ref.~\cite{Alday:2009zm}.

\subsection{MB representation for the tennis court diagram}
The procedure above can be repeated for the tennis court diagram
(fig.~\ref{fig-threeloop}b), which is given by
\be
\label{deftenniscourt}
I_{3b} = s \, t^2 \,  \int \frac{d^{4}x_{5} d^{4}x_{6} d^{4}x_{7}}{( i \pi^2)^3 } \, P_{17,m} \, \left( P_{25,m}\, P_{15,m}\,P_{57}\,P_{56}\,P_{16,m}\,P_{46,m} \,P_{67} \,P_{27,m}\,P_{37,m}  \,P_{47,m} \right)^{-1}\,.
\ee
A new feature of the tennis court is the presence of a numerator required by dual conformal symmetry.
It is possible to treat this numerator as a propagator with negative power. However, this would require an analytical continuation,
for example in the dimension of the integral, $D \to 4 -2\epsilon$. Since our integral is finite in four dimensions, we find it 
preferable to perform the calculation in such a way that all steps can be done in four dimensions.
This can be achieved by carrying out the two-loop ladder subintegral first (integrations over $x_5$ and $x_6$).
The numerator will then combine in a natural way with a propagator obtained from this subintegral, allowing us
to use a simple formula for the final integration over $x_{7}$.

As was explained above, we start by computing the subintegral $ I_{3b}^{(1)}$ defined by
\be
I_{3b}^{(1)} = \int \frac{d^{4}x_{5}}{ i \pi^2 }   \left( P_{25,m}\, P_{15,m}\,P_{57}\,P_{56} \right)^{-1} 
\ee
where
\be
I_{3b} = s \, t^2 \,  \int  \frac{d^{4}x_{6} d^{4}x_{7}}{( i \pi^2)^2 }  \, P_{17,m} \, I_{3b}^{(1)} \, \left( P_{16,m}\,P_{46,m} \,P_{67}\,P_{27,m}\,P_{37,m}  \,P_{47,m} \right)^{-1} \,.
\ee
It can be written as
\begin{equation}
 I_{3b}^{(1)} = \int_{0}^{\infty} \frac{ d\alpha_{i} \, \delta( \sum \alpha_1 + \alpha_2 -1 ) } {[ (P_{17,m} \alpha_1 + P_{27,m} \alpha_2 ) \alpha_7 + (P_{16,m} \alpha_1 + P_{26,m} \alpha_2 ) \alpha_6 + P_{67} \alpha_6 \alpha_7 + m^2  ]^2} \,.
\end{equation}
The integrals over $\alpha_6$ and $\alpha_7$ are done using \eqn{aux-4}.
Introducing two further MB integrals to factorize both $b$ and $c$  in \eqn{aux-4}, we arrive at
\be
\label{tennissub1}
 I_{3b}^{(1)} = \int \frac{dz_{1,2,3}}{(2 \pi i)^3} \, (m^2)^{-1-z_{1}} \,
 f_{3b}^{(1)}
\, (P_{16,m})^{z_1 - z_3 }\, (P_{27,m})^{z_1 - z_2 }\, ( P_{17,m})^{z_2 } \, (P_{26,m})^{z_3 } \, (P_{67})^{-1-z_{1}}
\ee
with
\begin{eqnarray}
f_{3b}^{(1)}
&=& \Gamma^2(1+z_1 )  \Gamma(-z_2 ) \Gamma( -z_3 ) \Gamma(-z_1 + z_2 ) \Gamma(-z_1 + z_3 ) \nonumber \\
&& \times  \frac{\Gamma(1+z_1+z_2 -z_3 ) \Gamma(1+z_1 -z_2 +z_3 ) }{\Gamma(2 (1+z_1 ))   }\,.
\end{eqnarray}
Next, we want to carry out the $x_{6}$ integration in \eqn{deftenniscourt}.
We define
\begin{equation}
I_{3b} \equiv 
s \, t^2 \, \int  \frac{d^{4}x_{7}}{i \pi^2} \, \frac{dz_{1,2,3}}{(2 \pi i)^{3}} \, f_{3b}^{(1)} \, (m^2)^{-1-z_{1}} \,  ( P_{17,m})^{1+ z_2 } \,   (P_{27,m})^{-1+z_1 - z_2 } \, ( P_{37,m})^{-1} \, ( P_{47,m})^{-1} \,  I_{3b}^{(2)}   \,.
\end{equation}
Collecting all propagators involving $x_{6}$ 
in \eqns{deftenniscourt}{tennissub1} we have
\begin{eqnarray}
 I_{3b}^{(2)} &=&  \int \frac{d^{4}x_{6}}{i \pi^2} \, (P_{67})^{-2-z_1 } \, (P_{16,m})^{-1+z_1 -z_3 } \, ( P_{26,m} )^{z_{3}} \, ( P_{46,m} )^{-1}  \\
 &=& \frac{1}{\Gamma(2+z_1 ) \Gamma(1-z_1 + z_3 ) \Gamma(-z_3 )} \int \frac{  d\alpha_i \, \delta(\sum_{1,2,4} \alpha_i -1 ) \, \alpha_1^{-z_1 + z_3 } \, \alpha_2^{-1-z_3 } \, \alpha_7^{1+z_1 } }
 {[t \alpha_2 \alpha_4 + (P_{17,m} \alpha_1 + P_{27,m} \alpha_2 + P_{47,m} \alpha_4) \alpha_7 + m^2 ]^2} \nonumber \\
&=& \frac{\Gamma(-z_1)}{\Gamma(1-z_1 +z_3 ) \Gamma( -z_3 )} \int_0^1 d\alpha_i \, \delta(\sum_{1,2,4} \alpha_i -1 ) \, \alpha_1^{-z_1 + z_3 } \, \alpha_2^{-1-z_3 } \, \times \nonumber \\
&& \times (t \alpha_2 \alpha_4 +m^2 )^{z_1 } \,   (P_{17,m} \alpha_1 + P_{27,m} \alpha_2 + P_{47,m} \alpha_4)^{-2-z_{1}}  \,.
\end{eqnarray}
This is almost identical to an integral considered earlier for the ladder diagrams.
Introducing MB parameters $z_{4,5,6}$ we find
\begin{eqnarray}
 I_{3b}^{(2)} =  \int \frac{dz_{4,5,6}}{(2 \pi i)^3} \, f_{3b}^{(2)} \, t^{z_{4}} \, (m^2)^{z_{1}-z_{4}} \, (P_{17,m})^{z_{5}} \, ( P_{27,m} )^{z_{6}}\,  ( P_{47,m} )^{-2-z_{1}-z_{5}-z_{6}} \,,
\end{eqnarray}
with
\begin{eqnarray}
f_{3b}^{(2)}  &=& \frac{ \Gamma(-z_{4}) \Gamma(-z_{1} + z_{4}) \Gamma(-z_{5})\Gamma(-z_6 ) \Gamma( 2 +z_1 +z_5 +z_6 )}{\Gamma(2+z_1 ) \Gamma(1-z_{1} +z_{3}) \Gamma(-z_{3})  \Gamma(2 (-z_{1} +z_{4}))} \, \times  \nonumber \\
&& \times \, \Gamma(1-z_{1} + z_{3} +z_{5} ) \Gamma(-z_{3} +z_{4} + z_{6} ) \Gamma(-1-z_{1} + z_{4} -z_{5} -z_{6} ) \,. 
\end{eqnarray}
Finally, we carry out the remaining integration over $x_{7}$,
\begin{eqnarray}
 I_{3b}^{(3)} \equiv   \int \frac{d^4x_{7}}{i \pi^2} \, (P_{17,m})^{1+z_{2}+z_{5}} \, (P_{27,m})^{-1+z_{1}-z_{2}+z_{6}} \, (P_{37,m})^{-1} \, (P_{47,m})^{-3-z_{1}-z_{5}-z_{6}} \,.
\end{eqnarray}
Up to a relabelling, this is the integral $I^{(2)}_{3a}$ of eqn (\ref{Iauxdef}).
Hence we have
\begin{eqnarray}
 I_{3b}^{(3)} 
 &=& \int \frac{dz_{7,8}}{ (2 \pi i)^{2}} \, f_{3b}^{(3)} \, s^{z_{7}} \, t^{z_{8}} \, (m^2)^{-2 -z_7 -z_8 } \,, 
\end{eqnarray}
with 
\begin{eqnarray}
f_{3b}^{(3)} &=& \frac{\Gamma(-z_{7}) \Gamma(-z_{8}) \Gamma(2+z_{7} + z_{8}) }{\Gamma(1-y_{1}) \Gamma(1-y_{3}) \Gamma(1+y_{1}+y_{3}) \Gamma(2 (2+z_{7}+z_{8})) } 
\times \nonumber \\
&& \times \, \Gamma(1+z_{7}) \Gamma(1+y_{1}+y_{3} +z_{7}) \Gamma(1-y_{1}+z_{8}) \Gamma(1-y_{3}+z_{8}) \,.
\end{eqnarray}
Here we have swapped $s$ and $t$ with respect to (\ref{Iauxfinal}) and 
$y_{1} \equiv -2 -z_{1}-z_{5}-z_{6}$ and 
$y_{3} \equiv z_{1}-z_{2}+z_{6}$.
Collecting all factors 
we 
obtain our final expression for the tennis court diagram
\begin{equation}
I_{3b} =  \int \frac{ dz_{1,2,3,4,5,6,7,8}}{(2 \pi i)^{8}} \,  \left(\frac{m^2}{s} \right)^{-1-z_{7}} \, \left(\frac{m^2}{t} \right)^{-2-z_{4}-z_{8}}   \, f_{3b}^{(1)}\, f_{3b}^{(2)}\, f_{3b}^{(3)}   \,.
\end{equation}
An allowed set of real parts for the integration variables is given by
\begin{equation}
z_{1} = -\frac{11}{16}, \; z_2 = -\frac{1}{2}, \; z_3 = -\frac{7}{16}, \; z_4 = -\frac{1}{8}, 
 \; z_5 = -\frac{9}{8}, \; z_6 = -\frac{3}{32}, \; z_7 = -\frac{13}{32}, \; z_8 = -\frac{1}{2}\,.
 \end{equation}
 
 \subsection{Auxiliary formulae}
Here we collect some auxiliary formulae that were used in the derivations
above.  It is understood that the formulae below should be used within
their region of validity, i.e. the real parts of the arguments of the
$\Gamma$ functions should be chosen positively.

\ba 
\label{MB-12}
(a+b)^{-\lambda} &=&  \frac{1}{\Gamma(\lambda)} \int \frac{dz_2}{(2 \pi i)} \Gamma(-z_{2}) \Gamma(\lambda+z_{2}) a^{z_{2}} b^{-\lambda -z_2 } \,,\\
(a+b+c)^{-\lambda} &=&  \frac{1}{\Gamma(\lambda)} \int \frac{ dz_1 dz_3 }{(2 \pi i)^2} \Gamma(-z_1 ) \Gamma(-z_3 ) \Gamma(\lambda + z_1 + z_3 )\, a^{z_1}\, b^{z_3}\, c^{-\lambda-z_1 -z_3} \,, \nonumber\\
&& \label{MB-13} \\
\label{alpha-int1}
\int_{0}^{\infty} dx\, x^{z_{1}} (a + b x)^{z_{2}} 
&=& a^{1+z_1 + z_2 } b^{-1-z_1 } \frac{\Gamma(1+z_1 ) \Gamma( -1-z_1 -z_2 )}{\Gamma(- z_2 )}\,, \\
\label{aux-4}
\int_0^{\infty} \frac{ dx \, dy}{ (a + b x + c y +d x y)^{2} } 
&=& (2 \pi i)^{-1} \int dz (a d)^z (b c)^{-1-z} \Gamma^2(-z) \Gamma^2(1+z)\,,
\\
\label{alpha-01}
\int_{0}^{1} \prod_{i=1}^{N} d\alpha_{i} \alpha_i^{q_i-1} \delta(1-\sum_{j=1}^{N} \alpha_{j}) 
&=& \frac{ \Gamma(q_1 ) \ldots \Gamma(q_N )}{\Gamma(q_1 + \ldots q_N )} \,.
\ea

\section{Numerical evaluation using sector decomposition}
\setcounter{equation}{0}

Here we present an alternative, semi-numerical approach for evaluating
the small $m^2$ expansion of integrals 
that is independent of the various MB representations derived above. 
We use the availability of powerful numerical algorithms for the calculation of the $\epsilon$-expansion of parameter 
integrals \cite{Binoth:2000ps,Bogner:2007cr}.
The expansion is done analytically, and only the coefficients of the poles
are evaluated numerically.  Therefore, this method avoids the problematic
large logarithms that would appear in a direct numerical integration.

In order to use these algorithms, we need to reformulate our problem 
as a calculation of the $\epsilon$-expansion of some parameter
integral.  
Consider a generic four-dimensional $L$-loop diagram
containing $a$ propagators, 
where propagators with indices $i \in {\mathcal M}$ have mass $m^2$, and all
other propagators are massless.\footnote{Modifications are necessary
for the case of integrals with numerator factors.} 
Then, the Feynman representation reads \cite{Smirnov:2006ry}
\begin{equation}
I =  \Gamma(a-2L) 
\, \int d\alpha_{i} \, \delta(1-\sum_{i} \alpha_{i})  \, U^{a-2L-2} 
\, \left[ V(s,t) + U m^2 \sum_{i \in {\mathcal M}}  \alpha_{i}  
\right]^{2 L - a}
\end{equation}
where $U$ and $V(s,t)$ are polynomials in the $\alpha_{i}$
(see ref.~\cite{Smirnov:2006ry} for more details).

We introduce one Mellin-Barnes parameter in order to separate off the mass dependence,
\begin{eqnarray}\label{mass-massless}
I &=& \int \frac{dz}{2 \pi i} \Gamma(-z) \Gamma(a-2L+z) 
\,  m^{2z}  \,f(s,t,z) \,, \\
f(s,t,z) &=& \int d\alpha_{i} \, \delta(1-\sum_{i} \alpha_{i}) 
\, ( \sum_{i \in {\mathcal M}}  \alpha_{i}  )^z   
\, U^{a-2L-2+z} 
\, V(s,t) ^{2L-a-z} \,,
\end{eqnarray}
with $2L-a<{\rm Re}(z)<0$.
Note that $f(s,t,z)$ is very similar to a dimensionally-regulated massless Feynman diagram of the same topology.\footnote{A. Zhiboedov has 
independently written  
down a similar formula at one loop which exhibits the relation between massive and massless Feynman integrals (private communication).}

Since we want to determine the behavior of $I$ as $m^2 \to 0$ 
it is convenient to shift the integration contour in $z$ to positive values of ${\rm Re}(z)$.
In doing so, one picks up a residue from the pole at $z=0$, and it is clear that the logarithms in $m^2$ will come from (minus) this residue. 
\begin{eqnarray}
I &=& - {\rm Residue}
\left[  \Gamma(-z) \Gamma(a-2L+z) m^{2z}  \,f(s,t,z) \right]\bigg|_{z=0}
 + O(m^2) \,.
\end{eqnarray}
The value of the latter can be obtained from the expansion of 
$\Gamma(-z)  \Gamma(a-2L+z) m^{2z} $  and $f(s,t,z)$ about  $z=0$.
Therefore, in order to compute $I$ up to and including order $\log^0(m^2)$, we need to evaluate $f(s,t,z)$ to order $z^0$.
Note that we do not need terms that vanish as $z \to 0$.
The coefficients of this expansion can be computed numerically 
using ref.~\cite{Bogner:2007cr}.

We have tested this method for the two-loop ladder and found agreement
with the result of ref.~\cite{Alday:2009zm}.  The method is also
applicable to three-loop integrals but requires more computer time
and memory.

\section{Regge limit of the $L$-loop ladder diagram}
\setcounter{equation}{0}
\label{sect-eden}

In the two- and three-loop amplitudes considered in this paper, 
we see that the leading log behavior in the Regge (b) limit 
(see discussion in the introduction and in 
sec.~\ref{sect-regge})
arises from the vertical ladder diagram.
In this appendix,
we will compute the first three leading terms in the Regge limit (b)
of the ladder integral,
employing an approach that was used in ref.~\cite{Eden}.

The $s \gg t$ limit of the vertical ladder is equivalent
to the $t \gg s$ limit of the horizontal ladder,
and this is what we will now examine.
The horizontal ladder diagram corresponds to the integral
\ba
\ILlad &=&  
s^L \, t \, \int \prod_{j=5}^{L+4} \left(
\frac{d^{4}x_{j}}{i \pi^2} \right)   \, ( P_{25,m} P_{15,m} P_{35,m})^{-1} 
\prod_{j=6}^{L+4}  ( P_{j-1,j} P_{1,j,m} P_{3,j,m}   )^{-1}
\, (P_{4,L+4,m} )^{-1}  
\ea
where all the internal propagators are massless while the propagators on
the periphery  of the diagram have mass $m$.
For each of the propagators, we use 
$(1/P) = i \int_0^\infty d\al ~ \exp(-i \al P)$
to obtain
\ba
\ILlad  &=&  
i^{3L+1} \, s^L \, t \, 
\intxyz 
\int \prod_{j=5}^{L+4} 
\left( \frac{d^{4}x_{j}}{i \pi^2} \right)  \times
\\
&&
\times \exp\left[
-i \al_0 P_{25,m} 
-i \sum_{\i=1}^{L-1} \al_\i P_{\i+4,\i+5} 
-i \al_L P_{4,L+4,m} 
-i \sum_{\j=1}^L  (\bet_\j P_{1,\j+4,m} + \ga_\j P_{3,\j+4,m}   )
\right] \,.
\nonumber
\ea
Recalling that $P_{ij,m} = x_{ij}^2 + m^2$,
we express the exponent as 
\be
-i \left(
{\bf{x^T}\cdot {\bf A} \cdot x} - 2~{\bf B^T \cdot x} + C + m^2 \sigma \right) 
\ee
where
$ {\bf x}  = (x_5, \cdots, x_{L+4})$.
We now integrate over ${\bf x}$  to obtain
\ba
\ILlad  &=&  
i^{L+1} \, s^L \, t \, 
\intxyz 
A^{-2} 
\exp\left[-i ( C - {\bf    B^T \cdot A^{-1} \cdot B} + m^2 \sigma ) \right] \,.
\ea
Setting $x_{12}^2=x_{23}^2 = x_{34}^2=x_{41}^2=0$,  one has
\be
C - {\bf    B^T \cdot A^{-1} \cdot B } = 
s {D_s\over A} +  t {D_t \over A},
\qquad
{\rm where~}s= x_{13}^2 
\quad {\rm and~} t=x_{24}^2 \,.
\ee
Putting everything together, we have 
\be
\ILlad  =  
i^{L+1} \, s^L \, t \, 
\intxyz 
A^{-2} 
\exp\left[-i t {D_t \over A} - i J \right],
\qquad
J = s{ D_s \over A} + m^2 \sigma 
\ee
where $A$, $D_s$, $D_t$, and $\sigma$ are polynomials
of degree $L$, $L+1$, $L+1$, and $1$ in the $\alpha$, $\beta$ and $\gamma$'s.
These polynomials may be constructed using graphical rules
\cite{Eden,Smirnov:2006ry}. 

To explore the $t \gg s$ limit of the horizontal ladder diagram,
it is useful to perform the Mellin transform \cite{Eden}
\ba
M (\eps) &=& \int_0^\infty  {d \tau \over \tau^\eps}  \ILlad  
\ea
with respect to $\tau = t/m^2$.
This gives
\ba
\label{Mellin}
M (\eps) 
&=&
\tau 
\, (i s)^L 
\, (i m^2)^{\eps}
\, \Gamma(1-\eps)  
\, F(\eps)
\ea
where
\ba
F(\eps)
&=& \intxyz 
D_t^{\eps-1}  
A^{-1-\eps}  \e^{-iJ}.
\ea
The large $\tau$ behavior of $\ILlad$
is determined by the behavior of $M(\eps)$ near $\eps=0$.
However,
$F(\eps)$ diverges at $\eps=0$,
since $ D_t = \prodk \al_i$,
so we need to do a Laurent expansion about $\eps=0$.
First we integrate by parts with respect to each of the $\al_i$
to obtain
\be
F(\eps)= { 1 \over \eps^{L+1} } 
\intmxyz
\left( \prodk \al_\i \right)^{\eps} 
{\partial^{L+1} \over \partial \al_0 \cdots \partial \al_L}
\left( \e^{-iJ} \over A^{1+\eps}  \right)
\ee
where we have taken $\eps > 0$ so that the surface terms at $\al_\i=0$ 
vanish.
Then writing
\be 
\left( \prodk \al_\i \right)^\eps = 
1
+
\left[ \sumi \al_i^\eps - (L+1) \right]
+
\left[ \sum_{i<j}  (\al_i\al_j)^\eps - L\sumi \al_i^\eps + {1\over 2}L(L+1) \right]
+\cO(\eps^3)
\ee
we obtain 
\be
\label{FK}
F(\eps) 
=
{1 \over \eps^{L+1} } 
\left\{ K + 
\left[ \sumi {K_i }  - (L+1) K \right]
+ 
\left[ \sum_{i<j}  K_{ij} - L \sumi {K_i}
 + {1\over 2} L(L+1) K \right]
+ \cO(\eps^3) 
\right\}
\ee
where we have performed (most of) the integrations over $\al_i$ 
\ba
K &=&
\intyz \left( \e^{-iJ} \over A^{1+\eps}  \right) \Bigg|_{\al_k =0}
\nonumber\\
K_i 
&=& 
\intyz
~d\al_i ~\eps\,\al_i^{\eps-1}
\left( \e^{-iJ} \over A^{1+\eps}  \right) \Bigg|_{\al_k =0, k \neq i}
\nonumber\\
K_{ij} 
&=& 
\intyz
~d\al_i ~d\al_j 
~\eps^2 \,\al_i^{\eps-1}\,\al_j^{\eps-1}
\left( \e^{-iJ} \over A^{1+\eps}  \right) \Bigg|_{\al_k =0, k \neq i,j} \,.
\ea
The integrals in this equation may be interpreted as arising from
(various deformations of) a horizontal $L$-loop ladder diagram 
in $d=2+2\eps$ dimensions \cite{Smirnov:2006ry}.
The rungs with $\al_k=0$ are contracted to a point
(and the corresponding propagator omitted), 
and the rungs with factors $\eps \al_i^{\eps-1}$
correspond to propagators raised to the power $\eps$.
Because most of the $\al_k$ are set to zero,
the $L$-loop diagram separates into a product 
of smaller diagrams, and the integrals factorize.
\begin{figure}
 \psfrag{A}[cc][cc]{$(a)$}
\psfrag{B}[cc][cc]{$(b)$}
\psfrag{s}[cc][cc]{$s$}
\psfrag{t}[cc][cc]{$t$}
\psfrag{dots}[cc][cc]{$\dots$}
 \centerline{
 {\epsfxsize15cm  \epsfbox{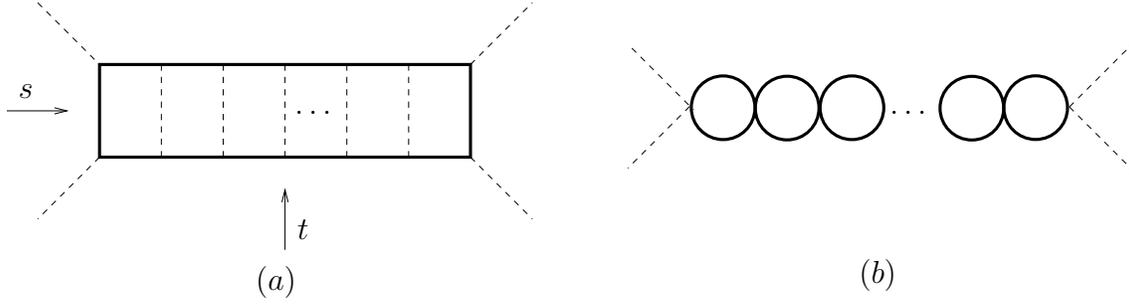}}
}
\caption{\small
Factorization of the LL order contribution to the 
Regge limit $t \gg s$ of the horizontal $L$-loop ladder integral (a) 
into $(2+2\epsilon)$-dimensional bubble integrals (b).}
\label{fig-regge1}
\end{figure}
For example, 
when all $\al_k=0$, we have
\ba
A
=
\prodl (\bet_\j + \ga_\j),
\qquad
{D_s\over A}  =
\sum_{\j=1}^L { \bet_\j \ga_\j \over \bet_\j + \ga_\j },
\qquad
\sigma =  \sum_{\j=1}^L  (\bet_\j + \ga_\j),
\ea
so 
$K$ separates into a product of one-loop bubble diagrams 
(cf. fig.~\ref{fig-regge1})
\be
K = \left[   (is)^{\eps-1}  B_1\right]^L
\ee
where the one-loop bubble is defined as 
\be
B_1 
\equiv  s^{1-\eps}
\int {d^{2+2\eps}  x_5 \over i \pi^{1+\eps} }
{1 \over P_{15,m} P_{35,m} }
= (is)^{1 - \eps}
\int_0^\infty  {d\bet d\ga \over (\bet+\ga)^{1+\eps} } 
\exp\left[-i s {\bet\ga \over \bet+\ga} -i  m^2 (\bet+\ga) \right] \,.
\ee
If we are only interested in the leading log (LL) behavior as $t \gg s$, 
we may set $\eps \to 0$, so that only the first term in \eqn{FK}
contributes, and $B_1$ becomes a two-dimensional diagram.   
This was noted on p. 134 of ref.~\cite{Eden}. 
(See also chapter 8 of ref.~\cite{Gribov}.)
Moreover, in this limit, the masses of the rungs 
of the original ladder diagram do not affect the result. 
Hence, in the LL limit, there is no distinction between the
Higgs regulator considered here and the cutoff regulator considered
in ref.~\cite{Schnitzer:2007rn}.

\begin{figure}
 \psfrag{t1}[cc][cc]{$T_{1}$}
\psfrag{t2}[cc][cc]{$T_{2}$}
\psfrag{b2}[cc][cc]{$B_{2}$}
\psfrag{b3}[cc][cc]{$B_{3}$}
\psfrag{eps}[cc][cc]{$\epsilon$}
 \centerline{
 {\epsfxsize15cm  \epsfbox{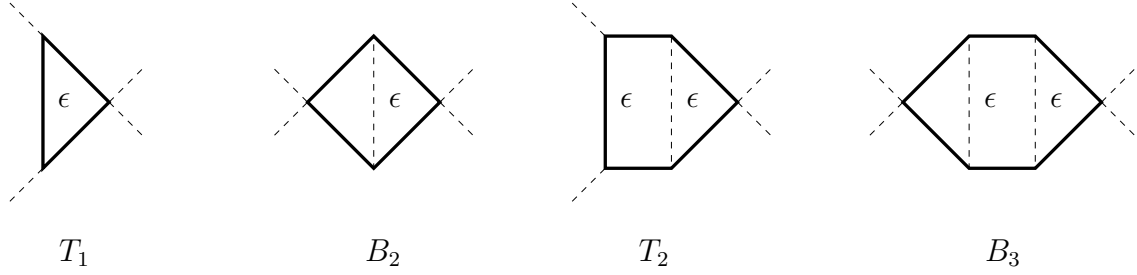}}
}
\caption{\small
Integrals arising in the computation of the Regge limit of the 
horizontal $L$-loop ladder integral to NNLL order.
The integration is in $2+2\epsilon$ dimensions.}
\label{fig-regge2}
\end{figure}
Taking into account the factorization of the integrals in \eqn{FK},
we obtain 
\ba
\label{F}
F(\eps) 
&=&
{\left(   i^{\eps-1}  B_1\right)^L
\over \eps^{L+1} }
\Big\{1 +  \big[   2 {t_1 } 
         + (L-1) {b_2 } - (L+1) \big]
+
\big[   
  	   2 {t_2} 
	+ {t_1 }^2
	+ 2(L-2) {t_1 b_2 } 
\nonumber\\
&&
	+ (L-2) {b_3 } 
	+ {1\over 2} (L-2) (L-3) {b_2^2 } 
	- 2L {t_1 } 
	- L(L-1) {b_2 } 
	+ {1\over 2} L (L+1)
\big]
\Big\}
\ea
with 
\ba 
\label{tb}
t_1 &=&  
\left(\Gamma(1+\eps)\over (is)^\eps\right)  
{T_1 \over B_1}, \qquad
b_2 = 
\left(\Gamma(1+\eps)\over (is)^\eps\right)  
{B_2 \over B^2_1} \nonumber\\
t_2 &=& 
\left(\Gamma(1+\eps)\over (is)^\eps\right)^2
{T_2 \over B^2_1}, \qquad
b_3 = 
\left(\Gamma(1+\eps)\over (is)^\eps\right)^2
{B_3 \over B^3_1} 
\ea
where $T_1$, $B_2$, $T_2$, and $B_3$ are
the $(2+2\eps)$-dimensional diagrams 
shown in fig.~\ref{fig-regge2},  and defined by
\ba 
T_1
&\equiv&  s
\int { d^{2+2\eps}  x_5 \over i \pi^{1+\eps} }
{ 1 \over P_{15,m} P_{35,m} (P_{25,m})^\eps  }
\nonumber\\
B_2
&\equiv& s^{2-\eps}
\int {d^{2+2\eps}  x_5 \over i \pi^{1+\eps} }
\int {d^{2+2\eps}  x_6 \over i \pi^{1+\eps} }
{1 
\over 
P_{15,m} P_{16,m} P_{35,m} P_{36,m} (P_{56})^\eps  }
\nonumber\\
T_2
&\equiv& s^2
\int {d^{2+2\eps}  x_5 \over i \pi^{1+\eps} }
\int {d^{2+2\eps}  x_6 \over i \pi^{1+\eps} }
{1 
\over 
P_{15,m} P_{16,m} P_{35,m} P_{36,m} (P_{25,m} P_{56})^\eps  }
\\
B_3
&\equiv& s^{3-\eps}
\int {d^{2+2\eps}  x_5 \over i \pi^{1+\eps} }
\int {d^{2+2\eps}  x_6 \over i \pi^{1+\eps} }
\int {d^{2+2\eps}  x_7 \over i \pi^{1+\eps} }
{1 
\over 
P_{15,m} P_{16,m} P_{17,m} 
P_{35,m} P_{36,m} P_{37,m}
(P_{56} P_{67})^\eps  } \,.
\nonumber
\ea
These integrals can be evaluated using MB techniques.

First consider the more general integral
\be
\int \frac{d^{d}x_{5}}{i \pi^{d/2}}  
(P_{13})^{a- \frac{d}{2}} \, 
(P_{15,m})^{-a_{1}} \, 
(P_{25,m})^{-a_{2}} \, 
(P_{35,m})^{-a_{3}} 
=   \int \frac{dz}{(2 \pi i)} \,  g_{1}(a_1,a_2,a_3;d;z)  \,  
u^{-z-a+\frac{d}{2}} 
\ee
where $a=a_1+a_2+a_3$ and $u = m^2/s$ and 
\begin{equation}
g_{1}(a_1,a_2,a_3 ; d; z) = \frac{ \Gamma(-z) \Gamma(a- \frac{d}{2} + z) \Gamma(a_1 + z) \Gamma(a_3 + z)} {\Gamma(a_1 ) \Gamma(a_3 ) \Gamma(2 z + a)} \,.
\end{equation}
We will also need
\begin{eqnarray}
&& \int  \frac{d^{d}x_{5}}{i \pi^{d/2}}  
(P_{13})^{a- \frac{d}{2}} \, 
(P_{15,m})^{-a_{1}} \, 
(P_{35,m})^{-a_{3}} \, 
(P_{65})^{-a_{6}}  \nonumber\\
&& \qquad 
=  \int \frac{dz_1\, dz_2\, dz_3}{(2 \pi i)^3} \,  
g_{2}(a_1, a_3, a_6; d; z_{i})  \, 
(P_{13})^{-z_1-z_3 } \,
(P_{16,m})^{z_1 } \,  
(P_{36,m})^{z_3 } \, 
u^{-z-a+\frac{d}{2}} 
\end{eqnarray}
where $a=a_1+a_3+a_6$
and $z=z_1+z_2+z_3$ and
\begin{eqnarray}
g_{2}(a_1 ,a_3 , a_6 ;d; z_i ) &=& 
\frac{ \Gamma(-z_{1}) \Gamma(-z_{2}) \Gamma(-z_{3}) } 
{ \Gamma(a_{1}) \Gamma(a_{3}) \Gamma(a_{6})} 
\Gamma(a-\frac{d}{2}+z) \times \\
&\times & \frac{ \Gamma(a_1 + z_1 + z_2 ) \Gamma(a_3 +z_2 +z_3 ) \Gamma( a_6 + z_1 +z_3 ) \Gamma(d-a-a_6 -z_1 -z_3 )}{\Gamma(a_1 + a_3 + z + z_2 ) \Gamma(d-a)} \,.
\nonumber
\end{eqnarray}
Then we may write
\begin{eqnarray}
B_{1}  
&=& \int \frac{dz}{(2 \pi i)} \, g_{1}(1,0,1;2+2 \epsilon;z) \, u^{-1+\epsilon-z} \,, \nonumber\\
T_{1} 
&=& \int \frac{dz}{(2 \pi i)} \, g_{1}(1,\epsilon,1;2+2 \epsilon ; z) \, u^{-1-z} \,, \nonumber\\
B_{2} 
&=& \int \frac{d^4z}{(2 \pi i)^4} \, g_{2}(1,1,\epsilon; 2+2\epsilon ; z_{1,2,3} ) \, g_{1}(1-z_{1},0,1-z_{3};2+2\epsilon;z_{4} ) \, u^{-2+\epsilon-z_{2}-z_{4}} \,,\nonumber\\
T_{2} 
&=& \int \frac{d^4z}{(2 \pi i)^4} \, g_{2}(1,1,\epsilon;2+2\epsilon;z_{1,2,3}) \, g_{1}(1-z_{1},\epsilon,1-z_{3};2+2\epsilon;z_{4}) \, u^{-2-z_{2}-z_{4}} \,,  
\nonumber\\
B_{3} 
&=& \int \frac{d^{7}z}{(2 \pi i)^7} \, 
g_{2}(1,1,\epsilon;2+2\epsilon;z_{1,2,3}) \, 
g_{2}(1-z_{1},1-z_{3}, \epsilon;2+2\epsilon; z_{4,5,6}) \, \nonumber \\
&& \times\quad 
g_{1}(1-z_{4},0,1-z_{6};2+2\epsilon; z_{7} ) u^{-3+\epsilon-z_{2}-z_{5}-z_{7}}\,. 
%B_{3} 
%&=& \int \frac{d^{7}z}{(2 \pi i)^7} \, 
%g_{2}(1,1,\epsilon;2+2\epsilon;z_{1,2,3}) \, 
%g_{2}(1,1,\epsilon;2+2\epsilon; z_{4,5,6}) \, \nonumber \\
%&& \times\quad 
%g_{1}(1-z_1-z_{4},0,1-z_3-z_{6};2+2\epsilon; z_{7} ) u^{-3+\epsilon-z_{2}-z_{5}-z_{7}}\,. 
\end{eqnarray}

Note that some of the MB representations above require an analytic continuation to $\epsilon \approx 0$. In some cases this may reduce the number of MB integrals.
Expanding around $\epsilon = 0$ 
(after having taken the analytic continuation where required) 
and then expanding in small $u$ we find\footnote{The coefficients of the
$\pi^4$ terms were obtained numerically,
and replaced by their probable rational equivalents.}
\begin{eqnarray}
\label{BT}
e^{-\epsilon \gamma_{\rm E}} \, B_{1} &=&  - 2 \log{u} + \epsilon \left[-
\log^2{u} - \frac{1}{3} \pi^2 \right]  + \epsilon^2 \left[ -\frac{1}{3}
\log^3{u} - \frac{1}{6} \pi^2 \log{u} + 4 \zeta_{3}  \right] + \ldots
\nonumber\\
e^{-\epsilon \gamma_{\rm E}} \, T_{1} &=&  - 2 \log u + \epsilon \left[ \frac{1}{2}  \log^2 u - \frac{2}{3} \pi^2 \right]  + \epsilon^2 \left[ \frac{1}{2} \pi^2  \log u +8  \zeta_{3}  \right] + \ldots
\nonumber\\
e^{-2 \epsilon \gamma_{\rm E}} \, B_{2} &=&  4 \log^2{u} + \epsilon \left[
\frac{10}{3} \log^3{u} + \frac{4}{3} \pi^2 \log{u}+ 4 \zeta_{3} \right]
\nonumber\\
&&+ \epsilon^2 \left[ \frac{5}{3} \log^4{u} + \pi^2 \log^2{u} + 4
\zeta_{3} \log{u} + \frac{2}{15} \pi^4 
\right] + \ldots
\\
e^{-2 \epsilon \gamma_{\rm E}} \, T_{2} &=&  4 \log^2 u + \epsilon \left[ \frac{1}{3} \log^3 u + 2 \pi^2 \log u + 4 \zeta_{3} \right] + \epsilon^2 \left[ \frac{5}{24} \log^4 u + \frac{31}{90} \pi^4 \right]+ \ldots
\nonumber\\
e^{-3 \epsilon \gamma_{\rm E}} \, B_{3} &=&  -8 \log^3 u + \epsilon \left[
-\frac{28}{3} \log^4 u - 4 \pi^2 \log^2 u - 16 \zeta_{3} \log u \right] + 
\nonumber\\ && +
\epsilon^2 \left[ -\frac{94}{15} \log^5 u - \frac{38}{9} \pi^2 \log^3 u - 20
\zeta_{3} \log^2 u - \frac{34}{45} \pi^4 \log u - 44.705 \right] + \ldots
\nonumber
\ea

Combining eqs.~(\ref{Mellin}), (\ref{F}), (\ref{tb}), and (\ref{BT}),
we obtain 
\ba
M(\eps) &=& 
\tau { (-2)^L \over \eps^{L+1} }
\Bigg\{ 1 
        + \eps   \left[ {L-1 \over 3}  \log^{L+1} u
       			+  {L+2 \over 6} \pi^2  \log^{L-1} u
       			+  (L-1) \zeta_3 \log^{L-2} u \right]
\nonumber \\
&+& 
         \eps^2 \bigg[ {10 L^2 - 14 L + 3 \over 180} \log^{L+2} u
	+	{(L+1)^2 \over 18} \pi^2 \log^{L} u
	+	{ L(L-2) \over 3} \zeta_3 \log^{L-1} u
\nonumber\\
&+&
		{(L+3)(5L+2) \over 360} \pi^4 \log^{L-2} u
	+ 	{L-2 \over 6} (L + 1.826) \pi^2 \zeta_3  \log^{L-3} u
\nonumber\\
&+&
		{(L-2)(L-3)\over 2} \zeta_3^2 \log^{L-4} u
\bigg]
+ \cO(\eps^3)
\Bigg\} \,.
\ea
To take the inverse Mellin transform, we use \cite{Eden}
\be
M(\eps) =  {1\over \eps^{n+1} }  \implies  I = {1 \over n!}
 {\log^{n} \tau \over \tau}
\ee
so that the first three leading log terms of the ladder diagram
are given by
\ba
&&
\hspace{-0.5in} 
\quad\mathrel{\mathop{\rm lim}\limits_{u,v \ll 1 }}
\;\; \mathrel{\mathop{\rm lim}\limits_{v \ll u}}\quad 
\ILlad  (u,v)
= 
  {2^L \over L!} \log^L v \log^L u 
\nonumber \\
&-&
 {2^L \over (L-1)!} \log^{L-1} v 
		 \left[ {1\over 3}  (L-1)  \log^{L+1} u
       			+  (L+2) \zeta_2   \log^{L-1} u
       			+  (L-1) \zeta_3 \log^{L-2} u \right]
\nonumber\\
&+&
 {2^L \over (L-2)!} \log^{L-2} v 
 	\bigg[ {10 L^2 - 14 L + 3 \over 180} \log^{L+2} u
	+	{(L+1)^2 \over 18} \pi^2 \log^{L} u
\nonumber\\
&+&		 { L(L-2) \over 3} \zeta_3 \log^{L-1} u
 	+	{(L+3)(5L+2) \over 360} \pi^4 \log^{L-2} u
	+ 	{L-2 \over 6} (L + 1.826) \pi^2 \zeta_3  \log^{L-3} u
\nonumber\\
&+&	{(L-2)(L-3)\over 2} \zeta_3^2 \log^{L-4} u
\bigg]
+ \cO( \log^{L-3} v )
\ea
where we recall that $v = m^2/t = 1/\tau$.
This result is used in sec.~\ref{sect-regge}.

\providecommand{\href}[2]{#2}\begingroup\raggedright

\endgroup

\begin{thebibliography}{99}

%\cite{Anastasiou:2003kj}
\bibitem{Anastasiou:2003kj}
  C.~Anastasiou, Z.~Bern, L.~J.~Dixon and D.~A.~Kosower,
  ``Planar amplitudes in maximally supersymmetric Yang-Mills theory,''
  Phys.\ Rev.\ Lett.\  {\bf 91} (2003) 251602
  [arXiv:hep-th/0309040].
  %%CITATION = PRLTA,91,251602;%%

%\cite{Bern:2005iz}
\bibitem{Bern:2005iz}
  Z.~Bern, L.~J.~Dixon and V.~A.~Smirnov,
  ``Iteration of planar amplitudes in maximally supersymmetric Yang-Mills
  theory at three loops and beyond,''
  Phys.\ Rev.\  D {\bf 72} (2005) 085001
  [arXiv:hep-th/0505205].
  %%CITATION = PHRVA,D72,085001;%%

%\cite{Drummond:2007cf}
\bibitem{Drummond:2007cf}
  J.~M.~Drummond, J.~Henn, G.~P.~Korchemsky and E.~Sokatchev,
  ``On planar gluon amplitudes/Wilson loops duality,''
  Nucl.\ Phys.\  B {\bf 795} (2008) 52
  [arXiv:0709.2368 [hep-th]].
  %%CITATION = NUPHA,B795,52;%%

%\cite{Alday:2007he}
\bibitem{Alday:2007he}
  L.~F.~Alday and J.~Maldacena,
  ``Comments on gluon scattering amplitudes via AdS/CFT,''
  JHEP {\bf 0711}, 068 (2007)
  [arXiv:0710.1060 [hep-th]].
  %%CITATION = JHEPA,0711,068;%%

%\cite{Bern:2008ap}
\bibitem{Bern:2008ap}
  Z.~Bern, L.~J.~Dixon, D.~A.~Kosower, R.~Roiban, M.~Spradlin, C.~Vergu and A.~Volovich,
  ``The Two-Loop Six-Gluon MHV Amplitude in Maximally Supersymmetric Yang-Mills
  Theory,''
  Phys.\ Rev.\  D {\bf 78} (2008) 045007
  [arXiv:0803.1465 [hep-th]].
  %%CITATION = PHRVA,D78,045007;%%

%\cite{Drummond:2008aq}
\bibitem{Drummond:2008aq}
  J.~M.~Drummond, J.~Henn, G.~P.~Korchemsky and E.~Sokatchev,
  ``Hexagon Wilson loop = six-gluon MHV amplitude,''
  Nucl.\ Phys.\  B {\bf 815}, 142 (2009)
  [arXiv:0803.1466 [hep-th]].
  %%CITATION = NUPHA,B815,142;%%

%\cite{DelDuca:2009au}
\bibitem{DelDuca:2009au}
  V.~Del Duca, C.~Duhr and V.~A.~Smirnov,
  ``An Analytic Result for the Two-Loop Hexagon Wilson Loop in  $\cN=4$ SYM,''
  arXiv:0911.5332 [hep-ph].
  %%CITATION = ARXIV:0911.5332;%%

%\cite{Drummond:2006rz}
\bibitem{Drummond:2006rz}
  J.~M.~Drummond, J.~Henn, V.~A.~Smirnov and E.~Sokatchev,
  ``Magic identities for conformal four-point integrals,''
  JHEP {\bf 0701} (2007) 064
  [arXiv:hep-th/0607160].
  %%CITATION = JHEPA,0701,064;%%

%\cite{Alday:2007hr}
\bibitem{Alday:2007hr}
  L.~F.~Alday and J.~M.~Maldacena,
  ``Gluon scattering amplitudes at strong coupling,''
  JHEP {\bf 0706} (2007) 064
  [arXiv:0705.0303 [hep-th]].
  %%CITATION = JHEPA,0706,064;%%

%\cite{Drummond:2007aua}
\bibitem{Drummond:2007aua}
  J.~M.~Drummond, G.~P.~Korchemsky and E.~Sokatchev,
  ``Conformal properties of four-gluon planar amplitudes and Wilson loops,''
  Nucl.\ Phys.\  B {\bf 795}, 385 (2008)
  [arXiv:0707.0243 [hep-th]].
  %%CITATION = NUPHA,B795,385;%%

%\cite{Brandhuber:2007yx}
\bibitem{Brandhuber:2007yx}
  A.~Brandhuber, P.~Heslop and G.~Travaglini,
  ``MHV Amplitudes in $\cN=4$ Super Yang-Mills and Wilson Loops,''
  Nucl.\ Phys.\  B {\bf 794} (2008) 231
  [arXiv:0707.1153 [hep-th]].
  %%CITATION = NUPHA,B794,231;%%

%\cite{Alday:2008yw}
\bibitem{Alday:2008yw}
  L.~F.~Alday and R.~Roiban,
  ``Scattering Amplitudes, Wilson Loops and the String/Gauge Theory
  Correspondence,''
  Phys.\ Rept.\  {\bf 468}, 153 (2008)
  [arXiv:0807.1889 [hep-th]].
  %%CITATION = PRPLC,468,153;%%

%\cite{Henn:2009bd}
\bibitem{Henn:2009bd}
  J.~M.~Henn,
  ``Duality between Wilson loops and gluon amplitudes,''
  Fortsch.\ Phys.\  {\bf 57}, 729 (2009)
  [arXiv:0903.0522 [hep-th]].
  %%CITATION = FPYKA,57,729;%%

%\cite{Drummond:2007au}
\bibitem{Drummond:2007au}
  J.~M.~Drummond, J.~Henn, G.~P.~Korchemsky and E.~Sokatchev,
  ``Conformal Ward identities for Wilson loops and a test of the duality with
  gluon amplitudes,''
  Nucl.\ Phys.\  B {\bf 826} (2010) 337
  [arXiv:0712.1223 [hep-th]].
  %%CITATION = NUPHA,B826,337;%%
  
%\cite{Drummond:2008vq}
\bibitem{Drummond:2008vq}
  J.~M.~Drummond, J.~Henn, G.~P.~Korchemsky and E.~Sokatchev,
  ``Dual superconformal symmetry of scattering amplitudes in $\cN=4$
  super-Yang-Mills theory,''
  Nucl.\ Phys.\  B {\bf 828}, 317 (2010)
  [arXiv:0807.1095 [hep-th]].
 %%CITATION = NUPHA,B828,317;%%

%\cite{Brandhuber:2008pf}
\bibitem{Brandhuber:2008pf}
  A.~Brandhuber, P.~Heslop and G.~Travaglini,
  ``A note on dual superconformal symmetry of the $\cN=4$ super Yang-Mills
  S-matrix,''
  Phys.\ Rev.\  D {\bf 78} (2008) 125005
  [arXiv:0807.4097 [hep-th]].
  %%CITATION = PHRVA,D78,125005;%%

%\cite{Drummond:2008cr}
\bibitem{Drummond:2008cr}
  J.~M.~Drummond and J.~M.~Henn,
  ``All tree-level amplitudes in $\cN=4$ SYM,''
  JHEP {\bf 0904} (2009) 018
  [arXiv:0808.2475 [hep-th]].
  %%CITATION = JHEPA,0904,018;%%

%\cite{Berkovits:2008ic}
\bibitem{Berkovits:2008ic}
  N.~Berkovits and J.~Maldacena,
  ``Fermionic T-Duality, Dual Superconformal Symmetry, and the Amplitude/Wilson
  Loop Connection,''
  JHEP {\bf 0809} (2008) 062
  [arXiv:0807.3196 [hep-th]].
  %%CITATION = JHEPA,0809,062;%%

%\cite{Beisert:2008iq}
\bibitem{Beisert:2008iq}
  N.~Beisert, R.~Ricci, A.~A.~Tseytlin and M.~Wolf,
  ``Dual Superconformal Symmetry from $AdS_5 \times S^5$ Superstring Integrability,''
  Phys.\ Rev.\  D {\bf 78} (2008) 126004
  [arXiv:0807.3228 [hep-th]].
  %%CITATION = PHRVA,D78,126004;%%

%\cite{Beisert:2009cs}
\bibitem{Beisert:2009cs}
  N.~Beisert,
  ``T-Duality, Dual Conformal Symmetry and Integrability for Strings on $AdS_5 x
  S^5$,''
  Fortsch.\ Phys.\  {\bf 57} (2009) 329
  [arXiv:0903.0609 [hep-th]].
  %%CITATION = FPYKA,57,329;%%

%\cite{Drummond:2009fd}
\bibitem{Drummond:2009fd}
  J.~M.~Drummond, J.~M.~Henn and J.~Plefka,
  ``Yangian symmetry of scattering amplitudes in $\cN=4$ super Yang-Mills theory,''
  JHEP {\bf 0905} (2009) 046
  [arXiv:0902.2987 [hep-th]].
  %%CITATION = JHEPA,0905,046;%%

%\cite{Korchemsky:2009hm}
\bibitem{Korchemsky:2009hm}
  G.~P.~Korchemsky and E.~Sokatchev,
  ``Symmetries and analytic properties of scattering amplitudes in $\cN=4$ SYM
  theory,''
  arXiv:0906.1737 [hep-th].
  %%CITATION = ARXIV:0906.1737;%%

%\cite{Bargheer:2009qu}
\bibitem{Bargheer:2009qu}
  T.~Bargheer, N.~Beisert, W.~Galleas, F.~Loebbert and T.~McLoughlin,
  ``Exacting $\cN=4$ Superconformal Symmetry,''
  JHEP {\bf 0911} (2009) 056
  [arXiv:0905.3738 [hep-th]].
  %%CITATION = JHEPA,0911,056;%%

%\cite{Sever:2009aa}
\bibitem{Sever:2009aa}
  A.~Sever and P.~Vieira,
  ``Symmetries of the $\cN=4$ SYM S-matrix,''
  arXiv:0908.2437 [hep-th].
  %%CITATION = ARXIV:0908.2437;%%

%\cite{Alday:2009zm}
\bibitem{Alday:2009zm}
  L.~F.~Alday, J.~M.~Henn, J.~Plefka and T.~Schuster,
  ``Scattering into the fifth dimension of $\cN=4$ super Yang-Mills,''
  arXiv:0908.0684 [hep-th].
  %%CITATION = ARXIV:0908.0684;%%

%\cite{Kawai:2007eg}
\bibitem{Kawai:2007eg}
  H.~Kawai and T.~Suyama,
  ``Some Implications of Perturbative Approach to AdS/CFT Correspondence,''
  Nucl.\ Phys.\  B {\bf 794} (2008) 1
  [arXiv:0708.2463 [hep-th]].
  %%CITATION = NUPHA,B794,1;%%

%\cite{Schabinger:2008ah}
\bibitem{Schabinger:2008ah}
  R.~M.~Schabinger,
  ``Scattering on the Moduli Space of $\cN=4$ Super Yang-Mills,''
  arXiv:0801.1542 [hep-th].
  %%CITATION = ARXIV:0801.1542;%%

%\cite{McGreevy:2008zy}
\bibitem{McGreevy:2008zy}
  J.~McGreevy and A.~Sever,
  ``Planar scattering amplitudes from Wilson loops,''
  JHEP {\bf 0808} (2008) 078
  [arXiv:0806.0668 [hep-th]].
  %%CITATION = JHEPA,0808,078;%%

%\cite{Mitov:2006xs}
\bibitem{Mitov:2006xs}
  A.~Mitov and S.~Moch,
  ``The singular behavior of massive QCD amplitudes,''
  JHEP {\bf 0705} (2007) 001
  [arXiv:hep-ph/0612149].
  %%CITATION = JHEPA,0705,001;%%

%\cite{Naculich:2007ub}
\bibitem{Naculich:2007ub}
  S.~G.~Naculich and H.~J.~Schnitzer,
  ``Regge behavior of gluon scattering amplitudes in $\cN=4$ SYM theory,''
  Nucl.\ Phys.\  B {\bf 794}, 189 (2008)
  [arXiv:0708.3069 [hep-th]].
  %%CITATION = NUPHA,B794,189;%%

%\cite{DelDuca:2008pj}
\bibitem{DelDuca:2008pj}
  V.~Del Duca and E.~W.~N.~Glover,
  ``Testing high-energy factorization beyond the next-to-leading-logarithmic
  accuracy,''
  JHEP {\bf 0805}, 056 (2008)
  [arXiv:0802.4445 [hep-th]].
  %%CITATION = JHEPA,0805,056;%%

%\cite{Naculich:2009cv}
\bibitem{Naculich:2009cv}
  S.~G.~Naculich and H.~J.~Schnitzer,
  ``IR divergences and Regge limits of subleading-color contributions to the
  four-gluon amplitude in $\cN=4$ SYM Theory,''
   JHEP {\bf 0910}, 048 (2009)
    [arXiv:0907.1895 [hep-th]].
 %%CITATION = JHEPA,0910,048;%%

\bibitem{DelDuca:2008jg}
  V.~Del Duca, C.~Duhr and E.~W.~N.~Glover,
  ``Iterated amplitudes in the high-energy limit,''
  JHEP {\bf 0812}, 097 (2008)
  [arXiv:0809.1822{\bf v4} [hep-th]].
  %%CITATION = JHEPA,0812,097;%%

\bibitem{DelDucaErr}
  V.~Del Duca, C.~Duhr and E.~W.~N.~Glover,
  ``Iterated amplitudes in the high-energy limit,''
  [arXiv:0809.1822{\bf v5} [hep-th]].

\bibitem{Eden}
  R.~J.~Eden, P.~V.~Landshoff, D.~I.~Olive, and J.~C.~Polkinghorne,
  {\it The Analytic S-Matrix} 
  (Cambridge University Press, Cambridge, 1966).
  
\bibitem{Brower:2008nm}
  R.~C.~Brower, H.~Nastase, H.~J.~Schnitzer and C.~I.~Tan,
  ``Implications of multi-Regge limits for the Bern-Dixon-Smirnov conjecture,''
  Nucl.\ Phys.\  B {\bf 814}, 293 (2009)
  [arXiv:0801.3891 [hep-th]].
  %%CITATION = NUPHA,B814,293;%%

\bibitem{Bartels:2008ce}
  J.~Bartels, L.~N.~Lipatov and A.~Sabio Vera,
  ``BFKL Pomeron, Reggeized gluons and Bern-Dixon-Smirnov amplitudes,''
  Phys.\ Rev.\  D {\bf 80}, 045002 (2009)
  [arXiv:0802.2065 [hep-th]].
  %%CITATION = PHRVA,D80,045002;%%

\bibitem{Bartels:2008sc}
  J.~Bartels, L.~N.~Lipatov and A.~Sabio Vera,
  ``$\cN=4$ supersymmetric Yang Mills scattering
  amplitudes at high energies: the Regge cut contribution,''
  arXiv:0807.0894 [hep-th].
  %%CITATION = ARXIV:0807.0894;%%

\bibitem{Brower:2008ia}
  R.~C.~Brower, H.~Nastase, H.~J.~Schnitzer and C.~I.~Tan,
  ``Analyticity for Multi-Regge Limits of the Bern-Dixon-Smirnov Amplitudes,''
  Nucl.\ Phys.\  B {\bf 822}, 301 (2009)
  [arXiv:0809.1632 [hep-th]].
  %%CITATION = NUPHA,B822,301;%%


%\cite{Schabinger:2009bb}
\bibitem{Schabinger:2009bb}
  R.~M.~Schabinger,
  ``The Imaginary Part of the $\cN = 4$ Super-Yang-Mills Two-Loop Six-Point MHV
  Amplitude in Multi-Regge Kinematics,''
  JHEP {\bf 0911} (2009) 108
  [arXiv:0910.3933 [hep-th]].
  %%CITATION = JHEPA,0911,108;%%

%\cite{Schnitzer:2007rn}
\bibitem{Schnitzer:2007rn}
  H.~J.~Schnitzer,
  ``Reggeization of $\cN=8$ Supergravity and $\cN=4$ Yang-Mills Theory II,''
  arXiv:0706.0917 [hep-th].
  %%CITATION = ARXIV:0706.0917;%%

%\cite{Mangano:1987xk}
\bibitem{Mangano:1987xk}
  M.~L.~Mangano, S.~J.~Parke and Z.~Xu,
  ``Duality and Multi-Gluon Scattering,''
  Nucl.\ Phys.\  B {\bf 298}, 653 (1988).
  %%CITATION = NUPHA,B298,653;%%

%\cite{Berends:1987cv}
\bibitem{Berends:1987cv}
  F.~A.~Berends and W.~Giele,
  ``The Six Gluon Process As An Example Of Weyl-Van Der Waerden Spinor
  Calculus,''
  Nucl.\ Phys.\  B {\bf 294}, 700 (1987).
  %%CITATION = NUPHA,B294,700;%%

%\cite{Mangano:1988kk}
\bibitem{Mangano:1988kk}
  M.~L.~Mangano,
  ``The Color Structure Of Gluon Emission,''
  Nucl.\ Phys.\  B {\bf 309}, 461 (1988).
  %%CITATION = NUPHA,B309,461;%%

%\cite{Broadhurst:1993ib}
\bibitem{Broadhurst:1993ib}
  D.~J.~Broadhurst,
  ``Summation of an infinite series of ladder diagrams,''
  Phys.\ Lett.\  B {\bf 307} (1993) 132.
  %%CITATION = PHLTA,B307,132;%%

%\cite{Bern:2006ew}
\bibitem{Bern:2006ew}
  Z.~Bern, M.~Czakon, L.~J.~Dixon, D.~A.~Kosower and V.~A.~Smirnov,
  ``The Four-Loop Planar Amplitude and Cusp Anomalous Dimension in Maximally
  Supersymmetric Yang-Mills Theory,''
  Phys.\ Rev.\  D {\bf 75} (2007) 085010
  [arXiv:hep-th/0610248].
  %%CITATION = PHRVA,D75,085010;%%

%\cite{Bern:2007ct}
\bibitem{Bern:2007ct}
  Z.~Bern, J.~J.~M.~Carrasco, H.~Johansson and D.~A.~Kosower,
  ``Maximally supersymmetric planar Yang-Mills amplitudes at five loops,''
  Phys.\ Rev.\  D {\bf 76} (2007) 125020
  [arXiv:0705.1864 [hep-th]].
  %%CITATION = PHRVA,D76,125020;%%

%\cite{Drummond:2008bq}
\bibitem{Drummond:2008bq}
  J.~M.~Drummond, J.~Henn, G.~P.~Korchemsky and E.~Sokatchev,
  ``Generalized unitarity for $\cN = 4$  super-amplitudes,''
  arXiv:0808.0491 [hep-th].
  %%CITATION = ARXIV:0808.0491;%%

%\cite{Brandhuber:2009xz}
\bibitem{Brandhuber:2009xz}
  A.~Brandhuber, P.~Heslop and G.~Travaglini,
  ``One-Loop Amplitudes in $\cN = 4$  Super Yang-Mills 
   and Anomalous Dual Conformal Symmetry,''
  JHEP {\bf 0908} (2009) 095
  [arXiv:0905.4377 [hep-th]].
  %%CITATION = JHEPA,0908,095;%%

%\cite{Elvang:2009ya}
\bibitem{Elvang:2009ya}
  H.~Elvang, D.~Z.~Freedman and M.~Kiermaier,
  ``Dual conformal symmetry of 1-loop NMHV amplitudes 
     in $\cN = 4$  SYM theory,''
  arXiv:0905.4379 [hep-th].
  %%CITATION = ARXIV:0905.4379;%%

%\cite{Brandhuber:2009kh}
\bibitem{Brandhuber:2009kh}
  A.~Brandhuber, P.~Heslop and G.~Travaglini,
  ``Proof of the Dual Conformal Anomaly of One-Loop 
   Amplitudes in $\cN = 4$  SYM,''
  JHEP {\bf 0910} (2009) 063
  [arXiv:0906.3552 [hep-th]].
  %%CITATION = JHEPA,0910,063;%%

%\cite{Korchemskaya:1992je}
\bibitem{Korchemskaya:1992je}
  I.~A.~Korchemskaya and G.~P.~Korchemsky,
  ``On lightlike Wilson loops,''
  Phys.\ Lett.\  B {\bf 287} (1992) 169.
  %%CITATION = PHLTA,B287,169;%%

%\cite{Smirnov:1999gc}
\bibitem{Smirnov:1999gc}
  V.~A.~Smirnov,
  ``Analytical result for dimensionally regularized massless on-shell  double
  box,''
  Phys.\ Lett.\  B {\bf 460}, 397 (1999)
  [arXiv:hep-ph/9905323].
  %%CITATION = PHLTA,B460,397;%%

%\cite{Nguyen:2007ya}
\bibitem{Nguyen:2007ya}
  D.~Nguyen, M.~Spradlin and A.~Volovich,
  ``New Dual Conformally Invariant Off-Shell Integrals,''
  Phys.\ Rev.\  D {\bf 77} (2008) 025018
  [arXiv:0709.4665 [hep-th]].
  %%CITATION = PHRVA,D77,025018;%%

%\cite{Bern:1994zx}
\bibitem{Bern:1994zx}
  Z.~Bern, L.~J.~Dixon, D.~C.~Dunbar and D.~A.~Kosower,
  ``One-Loop $n$-Point Gauge Theory Amplitudes, Unitarity and Collinear Limits,''
  Nucl.\ Phys.\  B {\bf 425}, 217 (1994)
  [arXiv:hep-ph/9403226].
  %%CITATION = NUPHA,B425,217;%%

%\cite{Bern:1994cg}
\bibitem{Bern:1994cg}
  Z.~Bern, L.~J.~Dixon, D.~C.~Dunbar and D.~A.~Kosower,
  ``Fusing gauge theory tree amplitudes into loop amplitudes,''
  Nucl.\ Phys.\  B {\bf 435}, 59 (1995)
  [arXiv:hep-ph/9409265].
  %%CITATION = NUPHA,B435,59;%%

%\cite{Buchbinder:2005wp}
\bibitem{Buchbinder:2005wp}
  E.~I.~Buchbinder and F.~Cachazo,
  ``Two-loop amplitudes of gluons and octa-cuts in $\cN = 4$ super Yang-Mills,''
  JHEP {\bf 0511} (2005) 036
  [arXiv:hep-th/0506126].
  %%CITATION = JHEPA,0511,036;%%

%\cite{Cachazo:2008dx}
\bibitem{Cachazo:2008dx}
  F.~Cachazo and D.~Skinner,
  ``On the structure of scattering amplitudes in $\cN=4$ super Yang-Mills and $\cN=8$
  supergravity,''
  arXiv:0801.4574 [hep-th].
  %%CITATION = ARXIV:0801.4574;%%

%\cite{Mandelstam:1965zz}
\bibitem{Mandelstam:1965zz}
  S.~Mandelstam,
  ``Non-Regge Terms in the Vector-Spinor Theory,''
  Phys.\ Rev.\  {\bf 137} (1965) B949.
  %%CITATION = PHRVA,137,B949;%%

%\cite{Grisaru:1973vw}
\bibitem{Grisaru:1973vw}
  M.~T.~Grisaru, H.~J.~Schnitzer and H.~S.~Tsao,
  ``Reggeization of yang-mills gauge mesons in theories with a spontaneously
  broken symmetry,''
  Phys.\ Rev.\ Lett.\  {\bf 30} (1973) 811;
  %%CITATION = PRLTA,30,811;%%

%\cite{Grisaru:1974cf}
\bibitem{Grisaru:1974cf}
  M.~T.~Grisaru, H.~J.~Schnitzer and H.~S.~Tsao,
  ``Reggeization of elementary particles in renormalizable gauge theories -
  vectors and spinors,''
  Phys.\ Rev.\  D {\bf 8}, 4498 (1973).
  %%CITATION = PHRVA,D8,4498;%%

%\cite{Grisaru:1979wi}
\bibitem{Grisaru:1979wi}
  M.~T.~Grisaru and H.~J.~Schnitzer,
  ``Reggeization Of Gauge Vector Mesons And Unified Theories,''
  Phys.\ Rev.\  D {\bf 20}, 784 (1979).
  %%CITATION = PHRVA,D20,784;%%

%\cite{Grisaru:1982bi}
\bibitem{Grisaru:1982bi}
  M.~T.~Grisaru and H.~J.~Schnitzer,
  ``Bound States In $\cN=8$ Supergravity And $\cN=4$ Supersymmetric Yang-Mills
  Theories,''
  Nucl.\ Phys.\  B {\bf 204} (1982) 267.
  %%CITATION = NUPHA,B204,267;%%

%\cite{Gluza:2007rt}
\bibitem{Gluza:2007rt}
  J.~Gluza, K.~Kajda and T.~Riemann,
  ``AMBRE - a Mathematica package for the construction of Mellin-Barnes
  representations for Feynman integrals,''
  Comput.\ Phys.\ Commun.\  {\bf 177} (2007) 879
  [arXiv:0704.2423 [hep-ph]].
  %%CITATION = CPHCB,177,879;%%

%\cite{Polyakov:1980ca}
\bibitem{Polyakov:1980ca}
  A.~M.~Polyakov,
  ``Gauge Fields As Rings Of Glue,''
  Nucl.\ Phys.\  B {\bf 164} (1980) 171.
  %%CITATION = NUPHA,B164,171;%%

%\cite{Korchemsky:1985xj}
\bibitem{Korchemsky:1985xj}
  G.~P.~Korchemsky and A.~V.~Radyushkin,
  ``Loop Space Formalism And Renormalization Group For The Infrared Asymptotics
  Of QCD,''
  Phys.\ Lett.\  B {\bf 171} (1986) 459.
  %%CITATION = PHLTA,B171,459;%%

%\cite{Ivanov:1985np}
\bibitem{Ivanov:1985np}
  S.~V.~Ivanov, G.~P.~Korchemsky and A.~V.~Radyushkin,
  ``Infrared Asymptotics Of Perturbative QCD: Contour Gauges,''
  Yad.\ Fiz.\  {\bf 44} (1986) 230
  [Sov.\ J.\ Nucl.\ Phys.\  {\bf 44} (1986) 145].
  %%CITATION = SJNCA,44,145;%%

%\cite{Korchemsky:1987wg}
\bibitem{Korchemsky:1987wg}
  G.~P.~Korchemsky and A.~V.~Radyushkin,
  ``Renormalization of the Wilson Loops Beyond the Leading Order,''
  Nucl.\ Phys.\  B {\bf 283} (1987) 342.
  %%CITATION = NUPHA,B283,342;%%

%\cite{Korchemsky:1991zp}
\bibitem{Korchemsky:1991zp}
  G.~P.~Korchemsky and A.~V.~Radyushkin,
  ``Infrared factorization, Wilson lines and the heavy quark limit,''
  Phys.\ Lett.\  B {\bf 279} (1992) 359
  [arXiv:hep-ph/9203222].
  %%CITATION = PHLTA,B279,359;%%

%\cite{Czakon:2005rk}
\bibitem{Czakon:2005rk}
  M.~Czakon,
  ``Automatized analytic continuation of Mellin-Barnes integrals,''
  Comput.\ Phys.\ Commun.\  {\bf 175} (2006) 559
  [arXiv:hep-ph/0511200].
  %%CITATION = CPHCB,175,559;%%

%\cite{Bogner:2007cr}
\bibitem{Bogner:2007cr}
  C.~Bogner and S.~Weinzierl,
  ``Resolution of singularities for multi-loop integrals,''
  Comput.\ Phys.\ Commun.\  {\bf 178} (2008) 596
  [arXiv:0709.4092 [hep-ph]].
  %%CITATION = CPHCB,178,596;%%

%\cite{Usyukina:1992jd}
\bibitem{Usyukina:1992jd}
  N.~I.~Usyukina and A.~I.~Davydychev,
  ``An Approach to the evaluation of three and four point ladder diagrams,''
  Phys.\ Lett.\  B {\bf 298}, 363 (1993).
  %%CITATION = PHLTA,B298,363;%%

%\cite{Smirnov:2006ry}
\bibitem{Smirnov:2006ry}
  V.~A.~Smirnov,
  {\it Feynman integral calculus}
  (Springer-Verlag, Berlin, 2006).

%\cite{Binoth:2000ps}
\bibitem{Binoth:2000ps}
  T.~Binoth and G.~Heinrich,
  ``An automatized algorithm to compute infrared divergent multi-loop
  integrals,''
  Nucl.\ Phys.\  B {\bf 585} (2000) 741
  [arXiv:hep-ph/0004013].
  %%CITATION = NUPHA,B585,741;%%

\bibitem{Gribov}
  V.~N.~Gribov,
  {\it The Theory of Complex Angular Momenta}
        (Cambridge University Press, Cambridge, 2003).

\end{thebibliography}
\end{document}